# Deciphering the Structure of Push-Pull Conjugated Polymer Aggregates in Solution and Film

Thomas P. Chaney[1], Christine LaPorte Mahajan[2], Masoud Ghasemi[2], Andrew J. Levin[1], Keith P. White[1], Scott T. Milner[2], Enrique D. Gomez[2,3], Michael F. Toney[1,4 *]

[1]*Materials Science and Engineering, University of Colorado, Boulder, CO 80309, USA*

[2]*Department of Chemical Engineering, The Pennsylvania State University, University Park, PA 16802, USA*

[3]*Department of Materials Science and Engineering, The Pennsylvania State University, University Park, PA, 16802, USA*

[4]*Department of Chemical and Biological Engineering and Renewable and Sustainable Energy Institute (RASEI), University of Colorado, Boulder, CO 80309, USA*

# 1. Abstract:

The morphology of conjugated polymer films is highly tunable, influencing their performance in organic electronics. Specifically, molecular packing or crystal structure strongly influence electronic processes such as light absorption and charge transfer. However, the unit cells of high-performance electron donor polymers remain unknown, limiting the understanding of how processing affects structure and device performance. This study characterizes the aggregate structure of PM6-type push-pull polymers using X-ray scattering, cryogenic electron microscopy, and molecular dynamics (MD) simulations. A novel forward simulation approach linking grazing-incidence wide-angle X-ray scattering (GIWAXS) with MD resolves a monoclinic unit cell that accurately describes PM6-type polymer aggregates in both thin films and casting solutions. Intimate π–π stacking between donor and acceptor units emerges from this unit cell. Analysis of experimental GIWAXS using this unit cell quantifies sliding disorder in these aggregates, which may impact device performance. The shape and internal structure of solution aggregates are also identified in chlorobenzene. These findings enhance our understanding of PM6-type polymer packing, outline a strategy for elucidating the crystal structure of weakly ordered materials, and provide an opportunity to control optoelectronic performance through aggregate formation in PM6 and other push-pull conjugated polymers.

# 2. Introduction:

Conjugated polymers have found use in a variety of electronics applications including organic photovoltaics (OPVs), biosensors, and thin film transistors. This is due to mechanical advantages (durability, flexibility) as well as their ability to form a network of interconnected domains through high aspect ratio aggregates and tie chains (*1*). Controlling this complex morphology is critical to optimizing organic electronic devices, but it is very difficult to characterize and hence control. Despite the large number of conjugated polymers reported to date, the crystal structure and chain packing motifs within crystals are only available for a few, select cases(*2*–*6*).

Grazing-incidence wide-angle X-ray scattering (GIWAXS) has been the tool of choice for the direct characterization of the aggregated morphology of conjugated polymer films. Specifically, analysis of the lamellar and π-π stacking peaks is commonplace when evaluating new chemistries, processing methods, or device degradation(*7*–*9*). Most GIWAXS analysis consists of fitting peak position in q-space, peak distribution in azimuthal (polar) angle chi, peak width, and integrated peak intensity. These values are then related to d-spacings between repeating structures in the aggregates, aggregate crystalline coherence lengths, and relative degrees of crystallinity(*10*). Ultimately the structure-function relationship drawn from these properties has been found to play a large role in the performance of organic electronic devices by governing charge transfer, recombination, and extraction(*11–14*).

Spectroscopic measurements such as UV-vis absorption and photoluminescence (PL) have additionally been used in literature to indirectly probe the morphological features of conjugated polymer films(*15–17*). In these measurements, characteristic peaks related to transitions between the ground and excited state are used to gain information related to the packing of polymers in aggregates. In the traditional H-J aggregate framework, the 0-0, or zero phonon, transition is allowed in chromophores that are packed in a head to tail arrangement due to the enhancement of the transition dipole moment(*18*). This type of packing is referred to as a J-aggregate. Conversely, chromophores packed in a head-to-head fashion cancel out the transition dipole moment, so the 0-0 transition becomes forbidden(*18*). This packing is referred to as an H-aggregate. Detailed analysis of the spectroscopic signals of these two types of aggregates has been championed by Frank Spano's research group(*19–21*). Unfortunately, the connection from H and J optical signature to structural aggregation becomes muddled in conjugated polymers as intra-chain interactions of a single polymer give rise to J-aggregate behavior due to the head to tail packing of chromophores(*19*). Additional complications to the H-J aggregate framework also arise in high performance conjugated polymers due to charge transfer state coupling, a topic that has recently been explored in literature(*15–17*). Thus, it remains challenging to connect packing structure and optical absorbance profiles of conjugated polymers without a direct structural probe.

Of recent interest to the organic optoelectronic community are high performance donor polymers constructed from monomers containing both an electron donating moiety and electron accepting moiety. These polymers are referred to as push-pull polymers and have led to smaller bandgaps and increased charge carrier mobility in many applications including OPVs (*22*, *23*). One particularly important push-pull polymer, PM6, has led to increasingly high-power conversion efficiencies reaching 19% when paired with Y6-type non-fullerene acceptors(*24–26*). PM6 is composed of two main units an electron donating benzo [1,2-b:4,5-b′] dithiophene (BDT) unit and an electron accepting 1,3-bis(thiophen-2-yl)−5,7-bis(2-ethylhexyl)benzo-[1,2-c:4,5-

c′]dithiophene-4,8-dione (BDD) unit shown in **Figure S1**. The recent success of the PM6 family of donor polymers has resulted in intense interest in molecular, solvent, and process engineering strategies to control and optimize its morphology for device performance. Both GIWAXS and optical spectroscopy have been extensively used to characterize PM6 films, yet no literature to our knowledge has clearly defined a unit cell. Importantly, we do not know if PM6 prefers to aggregate in a H-type head-to-head arrangement or rather a slip stack J-type arrangement along the π-stacking direction. The difference in this packing can significantly impact every step of power conversion in organic photovoltaics from light absorption to charge transport. Knowing the stacking arrangement of PM6 type polymers, and how it evolves during processing will provide insight into their high performance and show future avenues for further optimization of organic electronic materials and devices.

An abundance of morphology studies using GIWAXS have been carried out on PM6 polymers (*17*, *25*, *27–38*). All these studies have utilized a simple orthorhombic unit cell to index the peaks seen in GIWAXS images. This orthorhombic unit cell implies head-to-head packing of polymers along the π-stacking direction lining up the BDT unit with the neighboring polymer's BDT unit and likewise BDD with neighboring BDD units. This unit cell adequately explains the highest intensity lamellar and π-π stacking features seen on most GIWAXS images; however, it fails to properly account for a peak at 0.65 $Å^{-1}$ observed in the vast majority of GIWAXS images from literature and in our own GIWAXS shown in **Figure 1a-b**. In most literature, the 0.65 $Å^{-1}$ peak is not mentioned as the intensity is fairly low(*25*, *27–33*). Some studies identify the peak as a 2$^{nd}$ order lamellar peak (200)(*34–36*). However, this cannot be the case because its q-magnitude deviates from the expected position of 0.6 $Å^{-1}$ (twice the lamellar scattering vector of ~0.30 $Å^{-1}$). More importantly, the 0.65 $Å^{-1}$ peak is always situated at the in-plane azimuthal position regardless of the lamella peak's azimuthal position as shown in **Figures 1a-b**. A few studies account for this behavior suggesting that the peak at 0.65 $Å^{-1}$ is a backbone type peak of (001) (*17*, *37*). While this hypothesis accounts for the peaks azimuthal position, it does not explain the mismatch between the q-position and known monomer length for PM6. As pointed out by Cheng et al., the q-position suggests a repeat spacing of ~9.7 Å, yet the monomer length from density functional theory is 19.5 Å (*38*). This could mean that the peak seen here is a (002), but it would be unclear why the (001) peak is not present. We note that many other high performing push-pull conjugated polymers such as PTQ10, PTB7-Th, and D18 exhibit similar GIWAXS features to the PM6 features discussed here, suggesting that they may share a common packing motif(*29*, *39*, *40*).

In this manuscript, we resolve the unit cell of PM6 as well as probe the formation of ordered domains in the solution deposition process. This complete picture of aggregation from solution to film will provide insight into how molecular design and processing variables can be used to tune aggregation for optoelectronic performance. We have identified 16 potential stacking motifs based on the GIWAXS exhibited in both face-on dominant and edge-on dominant spun coat samples. Through a combination of molecular dynamics and a newly developed GIWAXS forward simulation tool, we identify a common slip-stack, traditionally J-aggregated stacking motif which leads to simulated X-ray scattering that agrees well with experimental GIWAXS. Importantly, we show that paracrystalline disorder, specifically slip disorder between π planes, is necessary to create the scattering profiles seen in experiment and provide a framework to quantify this slip disorder from experimental scattering patterns. Finally, we use a combination of small angle X-ray scattering and cryogenic electron microscopy (cryo-EM) to show paracrystalline aggregates exhibiting this slip stack motif readily form in solution over a large range in sizes. Together, these results reveal an experimentally verified PM6 packing structure and provide new insight into the formation of aggregates in solution. The conclusions provide valuable information necessary to better understand charge generation and transport within PM6, and to modify its aggregation to tune the optoelectronic properties. Furthermore, our analysis framework may be easily adapted to reveal packing motifs in a wide range of organic systems and other materials that exhibit paracrystalline disorder, such as water filtration films or hybrid perovskites.

# 3. Experimental Results and Discussion:

### 3.1 Experimental GIWAXS

The inconsistencies found in GIWAXS peak indexing of PM6 type polymers show that the simple orthorhombic unit cell alone does not adequately describe the aggregate morphology. To guide our search for the correct unit cell, GIWAXS was performed on films of PM6 deposited from two different solutions: a chloroform solution with 5 volume % 1-chloronaphthalene (CN) and a neat chloroform solution. The q-space converted detector images from the GIWAXS are shown **Figures 1 and S2**. While there are several impacts of the solvent additive CN, one well-documented impact is to slow the film drying process(*41*). The CN additive creates a film with more face-on oriented aggregates compared to the faster drying neat chloroform solution, which produces a film with more edge-on orientated aggregates. This is quantified in PM6 films through the corrected azimuthal intensity distribution of the lamella peak known as a pole figure which is given in the supplement **Figure S3** (*42*). Additionally, fitted peak centers and widths for the lamella, π-π, and backbone-type peaks are given in **Table S4**.

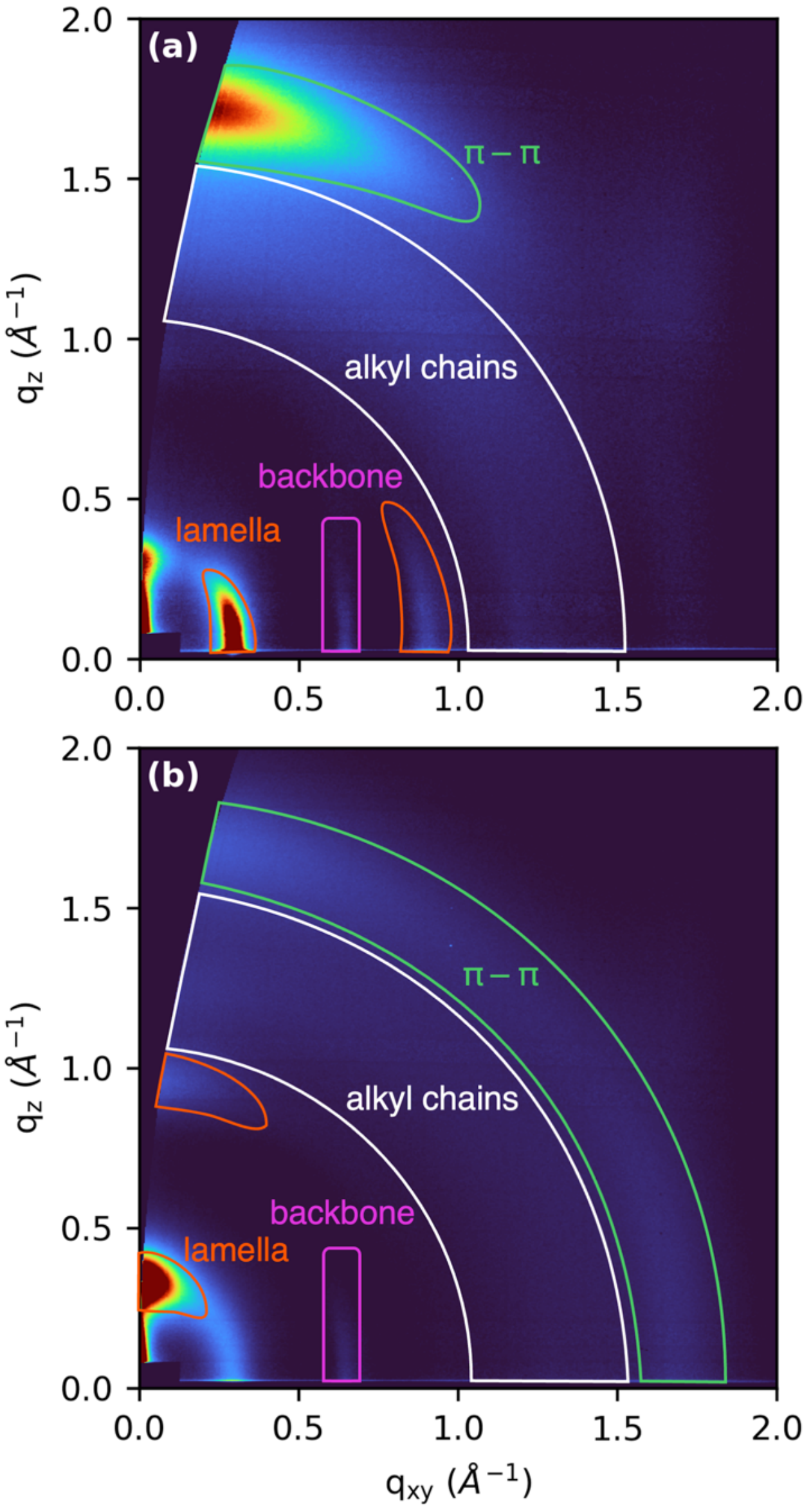

**Figure 1:** Linearly scaled GIWAXS detector images converted to q-space are shown for PM6 films deposited from (a) chloroform + 5 vol% CN and (b) neat chloroform. Scattering features are outlined for emphasis.

The ability to orient the same polymer aggregates in two different directions provides information on the orientational relationships between the observed scattering peaks that help discern what lattice spacings they may arise from. The details of assigning GIWAXS peaks to crystallographic planes are outside of the scope of this manuscript and are described in detail by Savikhin et al(*43*). At first inspection the GIWAXS images in **Figures 1 and S2** clearly show a lamella peak at ~0.3 $Å^{-1}$ and a π-π stacking peak at ~1.75 $Å^{-1}$. Additionally, weaker features are also resolved at ~1.3 $Å^{-1}$, ~0.9 $Å^{-1}$, ~0.65 $Å^{-1}$. The diffuse ring centered about 1.3 $Å^{-1}$ has been linked to alkyl chain packing in conjugated polymers(*44*). The 0.9 $Å^{-1}$ peak aligns well in both magnitude and azimuthal position to be a 3rd order lamella peak. Surprisingly, we note the absence of a 2nd order lamella peak. The cause for this absence of a 2nd order lamella peak is discussed later.

In the face-on aggregates (**Figure 1a**), we see the 0.65 $Å^{-1}$ peak oriented in-plane along with the lamella peak. In this case, it may seem reasonable to assign this peak as a 2nd order lamella peak. However, the edge-on oriented film (**Figure 1b**) GIWAXS reveals that the 0.65 $Å^{-1}$ peak remains along the $q_{xy}$ plane despite the new position of the lamella peak along the $q_z$ axis. This means that the 0.65 $Å^{-1}$ peak and the lamella peak must be orthogonal to each other and only appeared parallel to each other in the face-on GIWAXS due to the radial isotropy in the substrate plane for spin cast films(*43*, *45*, *46*). As discussed in the introduction, the consistent in-plane orientation of the 0.65 $Å^{-1}$ peak indicates it is likely a backbone-type peak. However, the standard orthogonal unit cell shows that the d-spacings between backbone planes would be the monomer length of 19.5 Å which does not agree with the 9.7 Å spacing indicated by the 0.65 $Å^{-1}$ peak. We also note that this 0.65 $Å^{-1}$ peak possesses a distribution in intensity primarily along the $q_z$ direction and not along the azimuthal arc like the other peaks. This suggests that the smearing seen for the 0.65 $Å^{-1}$ peak is due less to orientational disorder and more to another type of disorder. The in-plane orientation of the 0.65 $Å^{-1}$ peak suggests that the backbone repeats are parallel to the substrate plane. These few and relatively broad scattering features alone are not enough to resolve a unit cell, so we turn to computational simulations to provide insight.

### 3.2 Computational Studies

To match the scattering peaks shown in **Figures 1a-b** to an accurate unit cell, it is useful to first consider the different possibilities for polymer stacking that could produce the scattering peaks found in experimental GIWAXS. The new unit cell must ensure continuity of the polymer backbone, aromatic stacking with spacings of ~3.6 Å, and lamellar stacking with spacings of ~21 Å. Additionally, we must account for the asymmetry of the BDD unit about the backbone axis. Within these constraints we identify four configurations of the polymer stacking shown schematically in **Figures S5 and S6**:

1. *cis* vs *trans* arrangement of the BDD units along the backbone
2. Parallel (*para*) vs *anti*-parallel (*anti*) stacking of polymers
3. Backbone sliding offset between neighboring monomers in the π-stack direction (π-offset)
4. Backbone sliding offset between neighboring monomers in the lamella direction (lamella offset)

Taking into consideration the relative stability of each potential stacking motif, we identified that some configurations would be energetically unfavorable. Simulations of PM6 with *cis* arrangement of BDD units in the unit cell are unstable (**Figure S7**). We also did not explore *anti*-parallel backbone stacking along the lamella direction, as this would have large density fluctuations between the lamella, which is generally unfavorable. While *cis* vs. *trans* and parallel vs *anti*-parallel configurations are binary, we acknowledge that offsetting monomers can represent a spectrum of potential configurations. For simplicity, we will only consider configurations that have either a half monomer offset, where neighboring BDT and BDD groups roughly align, or no offset for our initial configurations of the simulations.

From these considerations, we have narrowed our search down to eight packing motifs shown schematically in **Figures S5 and S8**. These motifs are defined by the configurations of π-offset, lamella offset, and parallel vs anti-parallel stacking along the π-direction. Using these motifs, we built eight unique PM6 configurations to evaluate the relative stability, and we used a forward scattering simulation tool to directly compare each configuration to experimental GIWAXS. Each of the eight structural configurations is comprised of PM6 dimers in the all-trans configuration, with four lamella repeats and sixteen π-stacked repeats. We note that in maintaining the trans-conformation, the half monomer length is roughly 10 Å, which is slightly larger than the expected length of 9.7 Å from GIWAXS analysis.

Each PM6 configuration went through a two-stage equilibration process in our simulations, described in the methods, to evaluate their relative stabilities and to simulate the disorder observed in experimental GIWAXS. First, the chains were relaxed, initially freezing some atom positions in the backbone to maintain the initial lamella and π -offset. This structure is referred to as relaxed. After this, all atoms could move freely, and the total energy of each configuration was sampled at 300K to evaluate relative stability. Then the chains were allowed to freely equilibrate through a simulated annealing at 500K to encourage chain motion and then a slow cooling to 300K. This final structure is referred to as equilibrated. Real space images of the as-built, relaxed, and equilibrated configurations are shown in **Figure S9**.

By examining the relative energy ($U-U_{min}$) of each configuration before equilibration as shown in **Figure 2**, we see it is generally more favorable to have a backbone offset between π-stacks. However, the relative energy is indifferent to offsets in the lamella direction. **Figure 2** also shows the relative energy of each configuration after equilibration which shows all structures are within 2 kT/monomer (where k is the Boltzmann constant, T is 300 Kelvin) of each other except for the configuration with neither π-offset no lamella-offset, which had breaks in the π-stacked alignment. Within 2 kT/monomer, the differences in energy are similar to typical energy fluctuations at room temperature and may be considered degenerate.

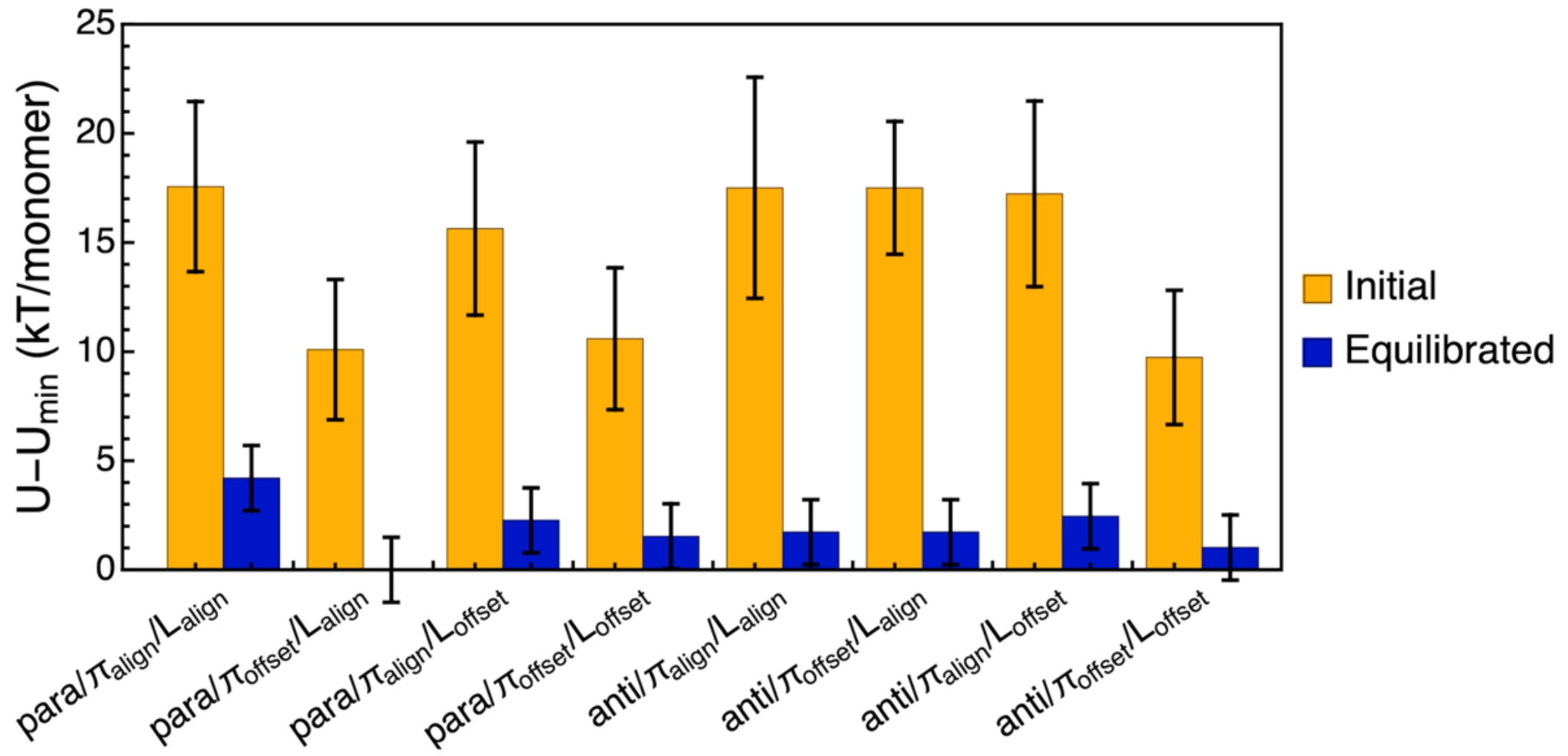

**Figure 2:** Energy difference of each configuration as compared to most stable configuration before and after equilibration. "Para" refers to parallel stacking of neighboring polymers along the π-π direction and "anti" refers to anti-parallel stacking.

Upon investigation of the equilibrated configurations from each starting configuration, we found that all equilibrated configurations contained some degree of π-offset, even if they are initialized without a π-offset. Additionally, no trend was observed regarding the lamella offset in equilibrated structures. We quantify the π-offset and lamella offset of equilibrated configurations in **Figures S10 and S11** respectively using probability distribution plots. Comparing the lamella distribution plots to the energy before and after equilibration, there does not appear to be a lamella offset that is more energetically favorable than others. In contrast, the probability distributions for the π-offset show all configurations rearrange to form motifs with some amount of π-offset, though the distribution after thermal annealing differs somewhat from a perfect half-monomer offset. These changes help decrease the energy of the system. We also show that polymer chains remain largely aligned along the π-stacking direction in **Figure S12**. Having identified energetically favorable stacking configurations, we carry out forward X-ray scattering simulations to compare the MD simulations with experimental GIWAXS.

### 3.3 Forward Scattering Simulations

To better understand the structural differences between the equilibrated structures as well as how they compare to experimentally probed film morphologies, we utilized a custom GIWAXS forward simulation tool(*47*). With this tool, we simulate GIWAXS detector images based on snapshots from each initial and equilibrated packing motif. To better approximate peak widths (through Scherrer broadening) of experimental face-on oriented PM6 GIWAXS, we propagated the 64-dimer configurations into larger rectangular slabs of varying dimensions. Slab dimensions were fit through comparison of simulated and experimental GIWAXS as described in the methods. The best fit sizes for each slab can be found in **Table S13**, with an average slab size of 180x58x94 Å along the respective real-space lamella, π-stack, and backbone directions. The simulated scattering of initial and equilibrated slabs can be found in **Figures S14 and S15**, respectively.

Interestingly, we find a striking similarity between the simulated scattering across all equilibrated configurations despite the differences shown in the initial configuration scattering. This suggests that all configurations have relaxed into similar stacking motifs as indicated by their equilibrated energies and π-offset distributions in **Figure 2 and S10**. Quantitative goodness of fit for each equilibrated configuration can be found in **Table S13**. Additionally, the similarity in simulated scattering patterns between equilibrated parallel vs *anti*-parallel stacked structures as observed in **Figure S15** indicates that experimental GIWAXS is insensitive to this stacking manipulation when significant disorder is present. We find the best fit from the equilibrated configurations is produced by an initially lamella offset structure that equilibrates to an approximately π-offset structure.

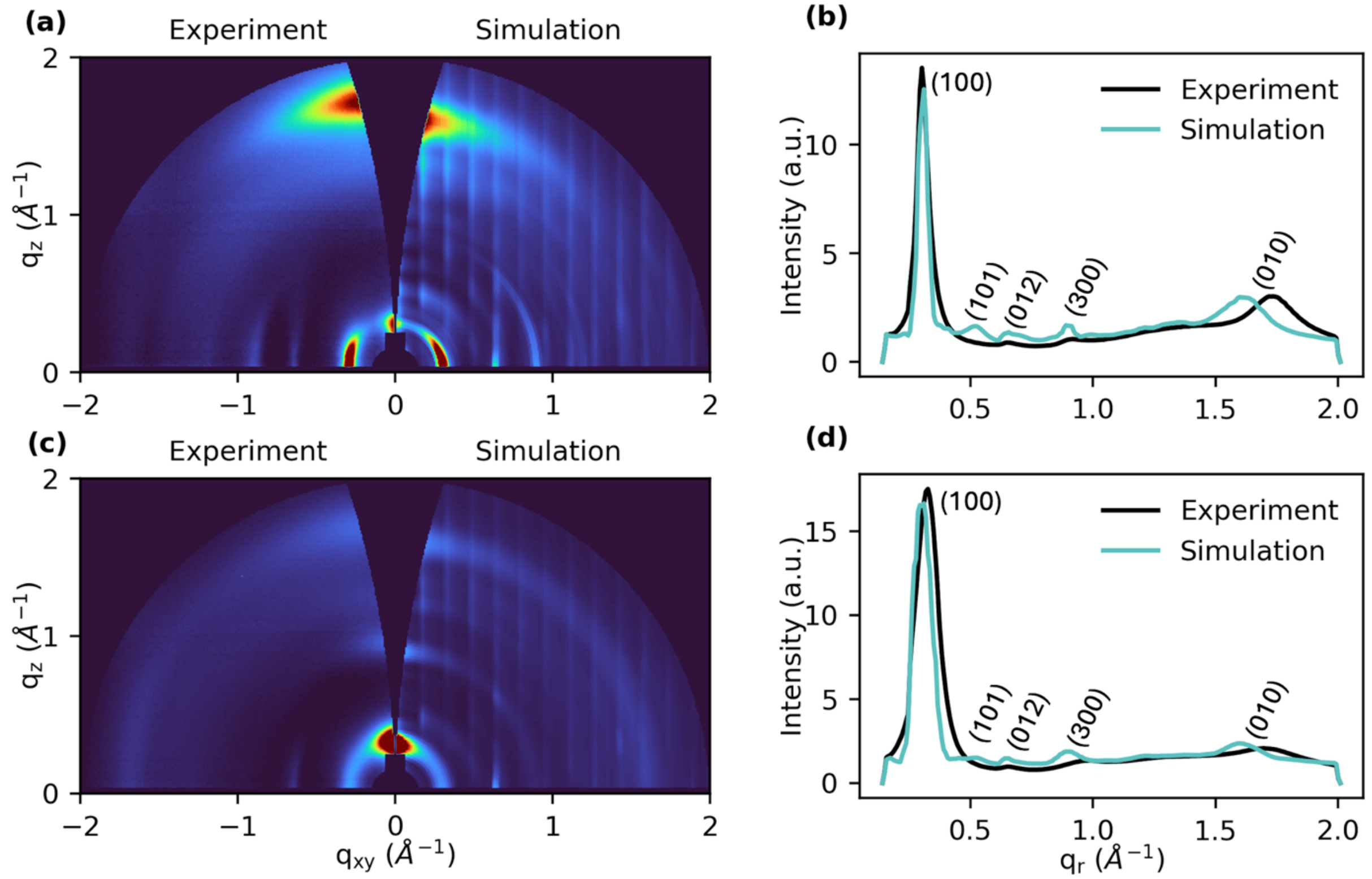

**Figure 3:** Comparison between experimental GIWAXS and forward simulated GIWAXS with a matched orientation distribution obtained from two different deposition solvents. Simulated GIWAXS shown here is from the extended equilibrated π-offset structure. (a) and (b) show 2D GIWAXS and 90-degree linecuts respectively for GIWAXS obtained from a film cast from CF + 0.5 vol% CN solution compared with simulation. (c) and (d) show 2D GIWAXS and 90-degree linecuts respectively for GIWAXS obtained from film cast from neat CF solution compared with simulation.

To better match the disorder in experimental films, particularly the $q_z$ distribution backbone type peak at $q_{xy} = 0.65$ Å$^{-1}$, we chose to carry out an extended equilibration of the π-offset configuration at a lower temperature. By equilibrating the relaxed π-offset structure for 2000 nanoseconds at 300 K, we find an even better agreement. The simulated GIWAXS from this extended equilibration structure is compared to experimental GIWAXS in **Figure 3**. While there is good agreement, there remain a few discrepancies between the simulated and experimental scattering data. First is in the position of the π-π (010) and lamella stacks (100), which is attributed to the force field used in the MD simulation not precisely representing chain packing in experimentally measured films. Second, is a faint peak at q=~0.5$Å^{-1}$ appearing just above the lamella peak in simulation. This peak is (101) according to the *trans* π-offset unit cell (**Figure S8**). Notably, this peak will change position for a *cis*- π-offset cell due to the shortening of the c-lattice vector. We hypothesize that frequent breaks in the trans-configuration of the PM6 backbone extinguish this peak in the experiment. We also note that when the log-scaled face-on oriented experimental GIWAXS (**Figure S2**) is closely examined, some faint intensity can be seen where the (101) peak appears in simulation. Finally, many periodic faint vertical streaks are seen in the simulation that have periodicity of 0.15 $Å^{-1}$ which are artifacts related to the 2-monomer length of the configuration box. Despite these discrepancies the agreement between experimental and simulated GIWAXS strongly suggests that this structure is representative of aggregates within experimentally produced films.

### 3.4 Proposed Unit Cell

Based on the above molecular dynamics and forward scattering results, we propose a generic π-offset unit cell (**Figure 4**) to describe PM6 aggregation. Dimensions for this unit cell were chosen based on the experimental GIWAXS peak positions. Note that the asymmetric BDD unit is now represented by a circle, signifying that this unit cell makes no claim as to the orientation of BDD units along or between backbones. This representation reflects the lack of a perfect *trans*- or *cis*-backbone configuration, as well as any self-consistent parallel or *anti*-parallel stacking. Additionally, this representation is further simplified by not depicting any lamella offset, as our MD simulations showed no preference for any specific lamella offset.

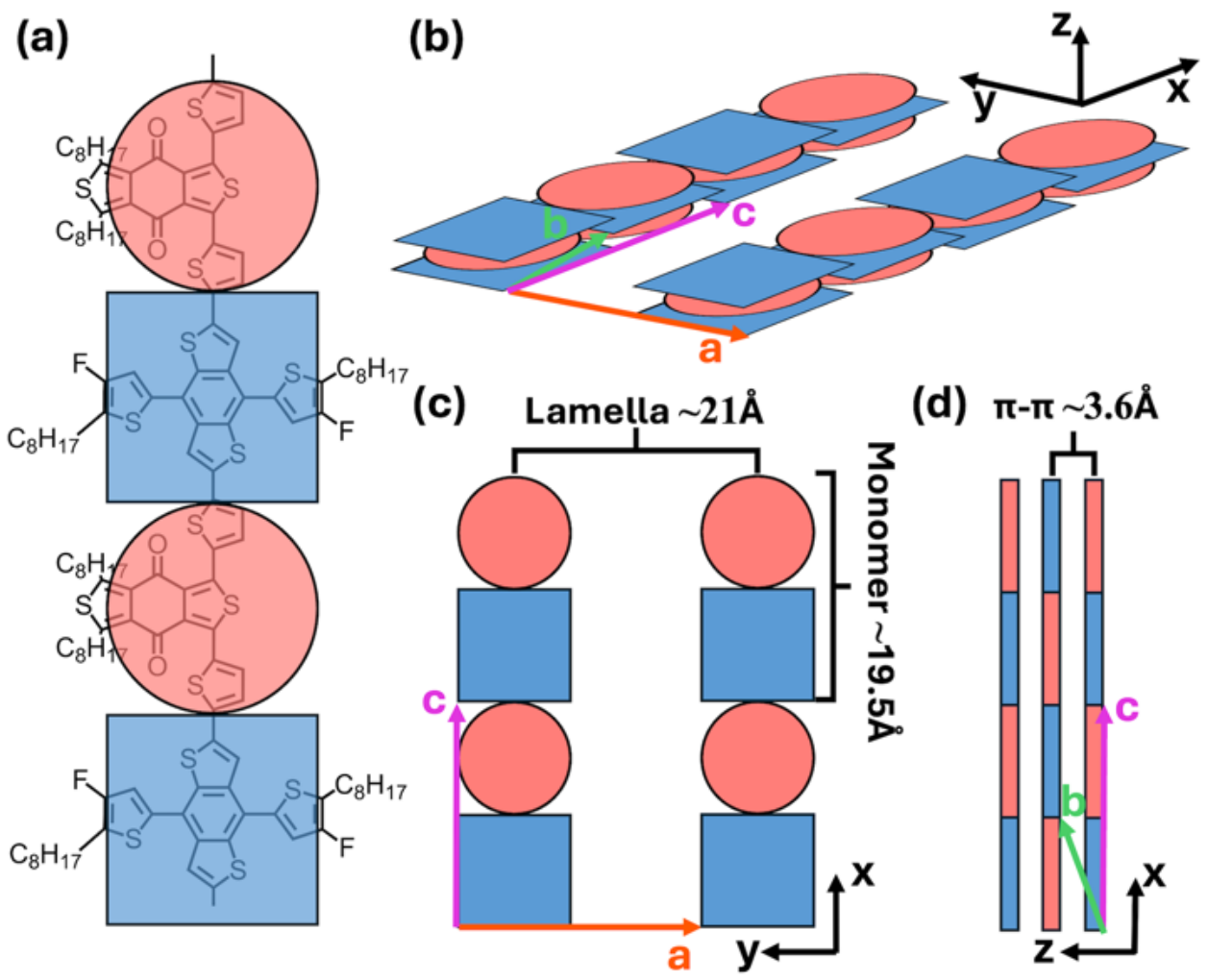

**Figure 4**: (a) schematic representation of PM6 aggregate structure. (b) schematic showing proposed "slip stack" monoclinic lattice with (c) top view and (d) side view.

To index observed GIWAXS peaks using the slip-stack unit cell, we convert the real-space unit cell from **Figure 4b** into reciprocal space shown in **Figure S16**. We can infer the respective azimuthal positions of two peaks from the relative angles between their associated reciprocal lattice vectors. In the newly proposed unit cell, the backbone peak is now indexed as (012), but this index still refers to the stacking planes along the polymer backbone, specifically at half monomer distances. As seen in **Figure S16**, the (100) and (012) planes are orthogonal, and their magnitudes are 0.30 $Å^{-1}$ and 0.65 $Å^{-1}$, respectively. This is in perfect agreement with the relative azimuthal position of experimental and simulated GIWAXS peaks shown in **Figure 3**. We note that the (001) peak using this unit cell is at a q-magnitude of 0.93 $Å^{-1}$ and would appear mostly out-of-plane for a face on orientation. While we do not observe the (001) peak in our experiment or simulated GIWAXS of equilibrated supercells, we do see it clearly in the initial configurations with π-offset (see Figure S14). Given the (001) planes in our unit cell stack primarily along the π-stacking direction, we attribute the absence of the (001) peak in both experimental and simulated GIWAXS patterns to paracrystalline disorder between π-stacked planes. Specifically, a distribution of π-offsets that will create a distribution of **b** lattice vectors with varying x-components results in a diminished and diffuse (001) peak that is not easily identified. We have seen wide distributions of π-offsets exist within the MD simulations of aggregates. The π-offset distribution from experimental GIWAXS is quantified in the following section. We also note the absence of a predicted (200) peak despite the presence of a (300) peak. This is an unusual feature in both the GIWAXS of experimental films, equilibrated supercells, and in literature GIWAXS of other conjugated polymer systems(*48*, *49*). In very disordered systems, including polymer aggregation, it is common to see higher ordered peaks extinguished(*50*). However, disorder alone cannot account for the lack of (200) when a (300) is present. Instead, the lack of a (200) peak is due to the extinction of even-ordered lamella peaks from a structure factor similar to that observed in some block copolymers (*51*, *52*). Particularly, the BDT moiety along the PM6 backbone exhibits conjugation in the lamella direction with a length of ~10.5 Å which is exactly half of lamellar d-spacing (21 Å). The resulting electron density profile along the lamellar direction is analogous to an infinite slit experiment with slits having the width exactly half of their center-to-center spacing. This arrangement results in forbidden even-order peaks (e.g., (200), (400), ...). A visualization of this effect and 1-dimensional simulation of scattering is included in **Figure S17**. In further support of this (200) peak extinguishing mechanism, we note of cases in literature where the lamellar distance for some BDT-containing polymers deviates from exactly two times the BDT width, and the (200) peak is no longer extinguished(*53–55*).

Through this reciprocal space unit cell analysis along with consideration of structure factor, we show that our proposed slip stack monoclinic lattice along with paracrystalline disorder adequately describes all peaks in the experimental GIWAXS profiles. Next, we analyze the unique shape of the (012) peak and how it relates to disorder in PM6 films.

### 3.5 Analysis of Paracrystalline Disorder

Knowledge of the unit cell enables a deeper analysis of disorder captured experimentally through GIWAXS of PM6 polymer films. To demonstrate this, we focus on the (012) peak. As discussed earlier the position of this peak is related to the reciprocal space lattice vectors **a***, **b***, and **c*** which are obtained via a transformation of the real space lattice vectors **a**, **b**, and **c**. We consider the scenario where in some aggregates the offset along the backbone direction for π-stacked polymer chains is not exactly half of a monomer (9.75 Å), but instead is equal another offset value, referred to as $\Delta_\pi$. As discussed earlier, disorder in $\Delta_\pi$ within aggregates may diminish the (001) peak as seen in the forward scattering simulations **Figures 3 and S15** Note that this scenario assumes homogenous *intra*-aggregate $\Delta_\pi$ values and a distribution of *inter*-aggregate $\Delta_\pi$ values within the scattering volume. In the case of a face-on oriented aggregate, the (012) peak occurs at the same $q_{xy}$=0.65 Å$^{-1}$ but now has a non-zero $q_z$ component given by equation 1:

$$q_{z_{(012)}} = \frac{-2\pi}{d_{(010)}} + \frac{4\pi\Delta_\pi}{d_{(010)} l_{monomer}} \qquad \text{eq. 1}$$

Where $d$ refers to d-spacing of the referenced peak and $l_{\text{monomer}}$ is the monomer length. In **Figure 1a** a strong vertical streak at $q_{xy}$=0.65 Å$^{-1}$ can be observed in agreement with this type of disorder. This disorder contrasts the orientational disorder commonly observed through azimuthal smearing in lamella and π-π peaks that corresponds to aggregate orientation distributions in the film.

Similarly, an analysis of offset along the backbone direction between neighboring lamella stacked polymer chains can be conducted. In our slip-stack unit cell (**Figure 4b**), this is normally 0 Å and here we call the offset $\Delta_{lamella}$. Analysis of the effect of this lamella slip disorder shows that $q_z$ remains at 0, but $q_{xy}$ varies according to equation 2:

$$q_{xy_{(012)}} = \sqrt{\left(\frac{-4\pi}{l_{monomer}}\right)^2 + \left(\frac{-4\pi\Delta_{lamella}}{d_{(100)} l_{monomer}}\right)^2} \qquad \text{eq. 2}$$

This shows that broadening of the (012) peak along $q_{xy}$ can be caused by a combination of Scherrer effects and paracrystalline disorder as described in literature(*50*), and also slip disorder between neighboring lamella. Thus, we are unable to link $q_{xy}$ peak broadening to any single type of disorder. The derivations for equations 1 and 2 for face-on geometry are given in the supplementary information.

Applying equation 1 to the (012) peak intensity integrated across $q_{xy}$ reveals a distribution of offset values for PM6 deposited from chloroform + 5 volume % CN, as shown in **Figure S18**. The most common offset value is seen at the half monomer length of 9.75 Å as expected, but there also exists a range of $\Delta_\pi$ values present in the film further confirming the presence of slip disorder.

**3.6 Investigation of Solution Aggregation**

We now move towards investigating how these aggregates form so we can better understand the relationship between deposition conditions and film morphology. To identify how early in the film formation process these ordered aggregates form along with their conformation, we utilized solution state small and wide-angle scattering (SWAXS) along with cryogenic electron microscopy (cryo-EM) on vitrified solutions. In these studies, we opted to use a chlorinated PM6 derivative, called PM7 (**Figure S1**), due to its higher electron density providing better scattering contrast for both SWAXS and cryo-EM measurements. PM7 is shown to possess the same slip stack aggregate structure as PM6 through the similarity in GIWAXS patterns of PM7 films shown in **Figure S2**. A chlorobenzene solvent was also chosen for these studies as it is another common deposition solvent for PM6 and PM7 and provides desirable properties (e.g. volatility and viscosity) for the vitrification process required for cryo-EM measurements.

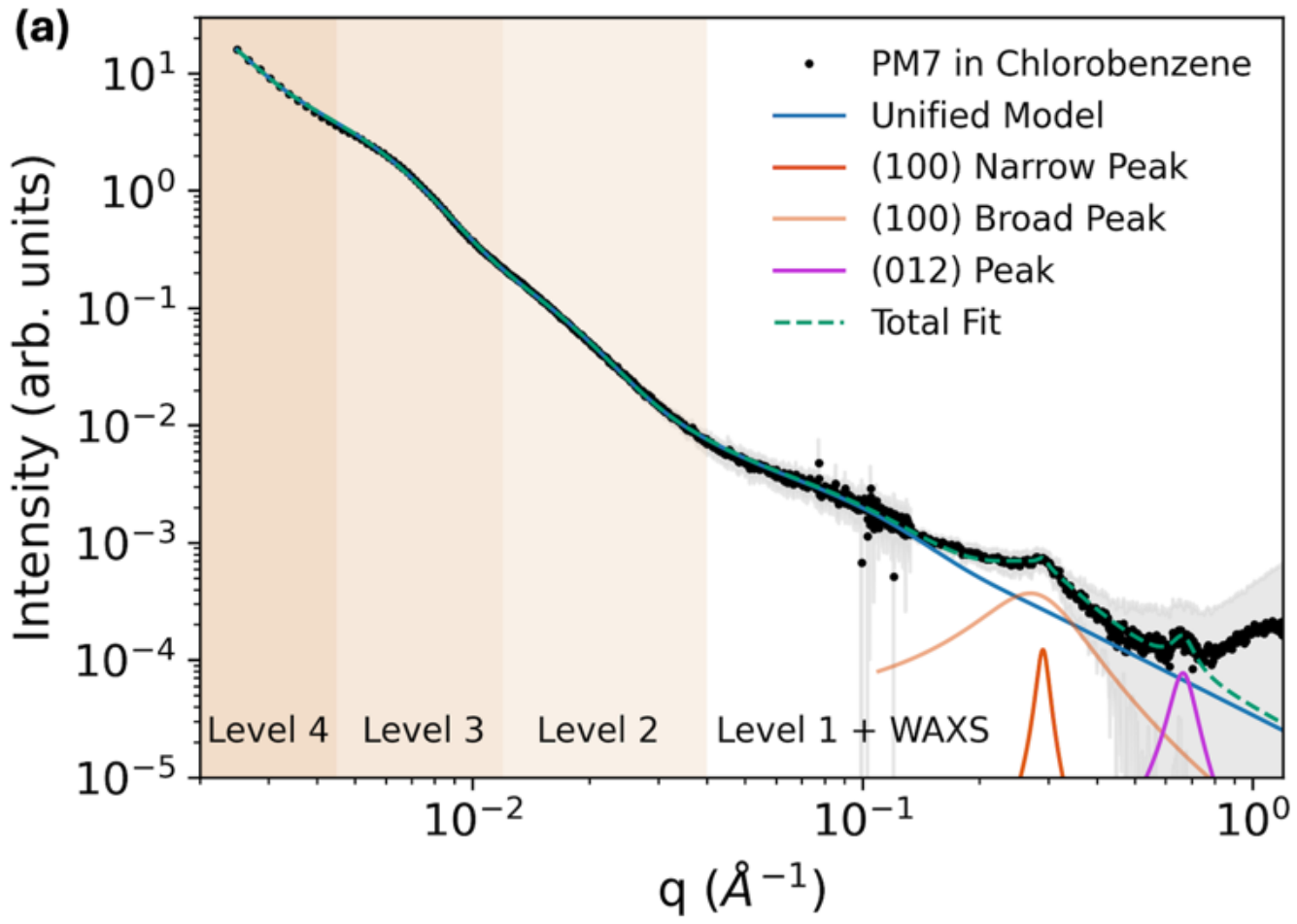

(b)

| Unified Model Fit Population | Rg (Å) | Porod Slope |
|---|---|---|
| Level 4 | >1000 | 2.9 |
| Level 3 | 360 | 3.8 |
| Level 2 | 130 | 3.1 |
| Level 1 | 20 | 1.7 |
| **Lorentzian Peak Fits** | **Center (Å$^{-1}$)** | **FWHM (Å$^{-1}$)** |
| (100) Narrow Peak | 0.29 | 0.02 |
| (100) Broad Peak | 0.27 | 0.08 |
| (012) Peak | 0.66 | 0.04 |

**Figure 5:** (a) Integrated and background subtracted scattering intensity profile of PM7 dissolved in chlorobenzene at 10mg/ml. Grey shading represents standard error of scattering intensity. A unified model with four levels, denoted by colored shading, was used to fit the small-angle scattering region. Lorentz peaks were used to fit features in the wide-angle scattering region(*56*). Selected fit parameters are shown in (b) $R_g$ is electron density radius of gyration and Porod Slope is defined as $P$ in the equation $I(q) = q^{-P}$ .

The azimuthally integrated solution SWAXS data is shown in **Figure 5**. In the high q regime, we see a broad scattering peak at 0.3 Å$^{-1}$ corresponding to (100) lamella stacking as well as a 0.66 Å$^{-1}$ peak that corresponds well to a (012) spacing within our slip stack unit cell. The presence of these peaks suggests that ordered aggregates form in the solution state before deposition has begun. The width of the (100) peak indicates that aggregates are forming with coherence lengths of 280 Å along the lamella direction. Given this large size we also assume that π-stacking is present in these solution aggregates although not observable in our q-range. In the low-q scattering there are several other features that contain information about the size and shape of these aggregates. While previous literature has utilized semi-flexible cylinder models to describe aggregation of similar polymers(*57*, *58*), these models cannot adequately fit this data, so a generalized Guinier-Porod model is used(*56*, *59*). Forward simulation approaches to fit this data such as those used in recent literature may provide more details but is outside the scope of this manuscript(*60*). From our tiered Guinier-Porod fit, the lowest-q shoulder corresponds to large aggregates with radius of gyration (Rg) ~380 Å, which are likely responsible for the scattering peaks in the high-q discussed earlier(*56*, *59*). Following that shoulder is a Porod slope of ~3 suggesting these aggregates are non-spherical(*59*, *61*). Additionally, there are two shoulders higher in q (levels 2 and 1 in **Figure 5**) corresponding to a population with an Rg of 140 Å which likely correspond to smaller aggregates along with a population with Rg of 20 Å which is attributed to disaggregated individual polymers.

In addition to solution SWAXS, we conducted cryo-EM on very thin layers of a vitreous solution (see the Methods section for more information). As shown in **Figure 6a**, cryo-EM reveals the presence of PM7 aggregates in the solution phase, with two distinct families of crystal planes corresponding to lamella stacking and backbone order. The aggregate shape for this example is a parallelogram with an internal angle of ~40°, which may suggest a similar angle between lamella stacked polymers. However, substantial variability in aggregate shape is apparent in additional cryo-EM images (**Figure S19**). As can be seen from the fast Fourier transform (FFT) of the cryo-EM micrograph (**Figure 6b-c**), spacings of 21 Å (q=0.3 $Å^{-1}$) and 9.5 Å (q=0.66 $Å^{-1}$) are apparent for (100), and (012), respectively, which is consistent with solution SWAXS and GIWAXS results. These results confirm the orthogonality of (100), and (012) planes, in line with the formation of the slip stacked unit cell in the solution phase.

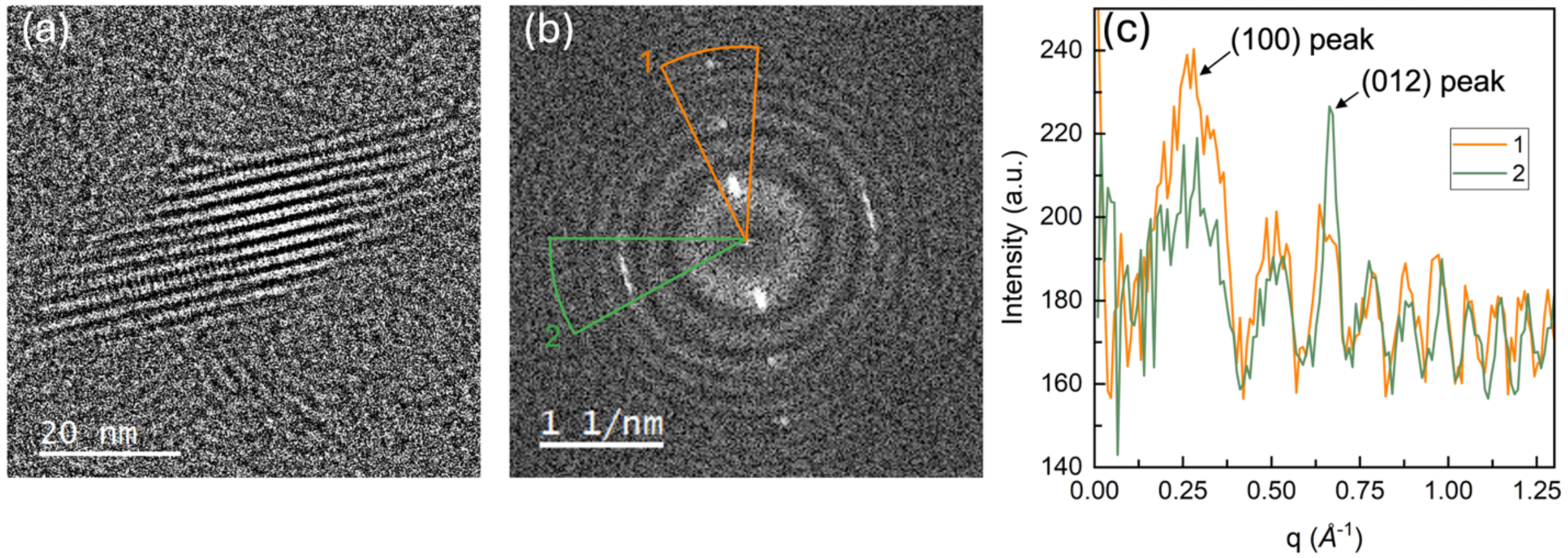

**Figure 6:** (a) Cryo-EM micrograph of vitrified PM7 in a chlorobenzene solution showing an isolated PM7 aggregate. (b) FFT of the cryo-EM micrograph in (a), and (c) two linecuts of the FFT showing orthogonal lamella (100) and backbone (012) reflections.

### 3.7 Conclusions

To conclude, the simple orthorhombic cell used widely for GIWAXS analysis of PM6-type polymers fails to account for a commonly seen 0.65 Å$^{-1}$ peak. Instead, a new slip-stacked unit cell, corroborated by MD, accounts for all peaks seen in GIWAXS. This new cell implies a J-type aggregation scheme where polymers are arranged head-to-tail along the π-stacking direction. The framework of combining MD and GIWAXS forward simulation has been shown to be a powerful tool to decipher GIWAXS patterns, especially for systems like conjugated polymers that are often dominated by paracrystalline disorder. We also demonstrate analysis of disorder from GIWAXS detector images beyond the standard pole figure analysis, providing rich insights into the aggregate structures within films.

Our investigations into the solution phase of these polymers reveals that well-ordered slip stack aggregates are present before deposition begins. Cryo-EM provides real-space images of these aggregates which are 100s of Ångstroms in diameter and exhibit the same ordering as seen in films. This shows that solvent pre-aggregation is a dominant factor in determining the final film morphology for PM6-type polymers.

Our solution characterization implies that focus should be given on how molecular engineering and polymer-solvent interactions, not only solvent vapor pressure, may impact aggregation which happens readily in the solution phase. Of specific interest is controlling the paracrystalline disorder that is found to be a large contributor to the scattering of PM6-type polymers, and likely also dominates the charge generation and transfer in optoelectronic devices. These results on PM6 and PM7 provide new opportunities for connection of morphology to spectroscopy, charge generation, and charge transfer within organic optoelectronics. Further, the framework demonstrated here is widely applicable to many thin film materials that possess significant orientational and paracrystalline disorder, such as hybrid perovskites and water filtration membranes.

# 4. Methods:

**GIWAXS**

PM6 and PM7 polymer used to make films were obtained from 1-material. Molecular weights of 100kDa and 90kDa with polydispersity 2.5 of were reported from gel permeation chromatography for PM6 and PM7 respectively. Chloroform and 1-chloronaphthalene solvents used were obtained from Sigma Aldrich. Films for GIWAXS were prepared by spin coating onto UV-ozone cleaned 10 x 10mm silicon substrates. First solutions were prepared by dissolving PM6 or PM7 polymers into either neat chloroform or a chloroform + 1-chloronaphthalene mixture at a concentration of 15mg polymer per milliliter of solvent. Solutions were stirred for >3 hours at 30°C and then 50uL was used to spin coat onto the substrates at 2000RPM for 40s. GIWAXS was taken at the National Synchrotron Light Source II on beamline 11-BM within an evacuated chamber to mitigate beam damage. Exposures of 20 seconds were taken at an incident angle of 0.15° using an X-ray beam with dimensions 50 x 200 um and an energy of 12.7 keV. The custom L-shaped Pilatus 800K detector was placed at a sample-to-detector distance of 259 mm which was verified with $CeO_2$ calibrant. Detector images were processed using the pyFAI python package (*62*).

**Molecular Dynamics:**

Molecular dynamics simulations were carried out in GROMACS using virtual site coarse grained forcefields described in prior work(*63*). Atom trajectories in these moieties are described by three atoms, while constraining bonds within the ring. Bonded forcefields are used to describe flexible bonds between virtual site moieties and the alkyl side-chains. In addition, hydrocarbon components are represented with united atom forcefields to further reduce the degrees of freedom in simulations. By creating bonded potentials between the first and last monomer in the chain backbone, the polymer was connected over the periodic boundary to make infinitely long chains without end effects.

Each PM6 configuration was built with 4 repeats in the lamella direction and 16 in the π-stacked direction. Before comparing the configurations to experiment, we first relax the chains to remove any unrealistically high forces from the building process. To avoid disturbing the offset configuration between chains while the system relaxes, we first perform an initial relaxation procedure in which the position of one atom within each backbone monomer was frozen, allowing free motion among all other atoms. Under this condition, we minimize the energy, followed by NVT simulation at 300 K for 60 ns. This was then followed by energy minimization, freezing atom positions to obtain energetically relaxed but idealized structures which could then be compared with experimental GIWAXS scattering. The energy for each relaxed structure was evaluated after simulating for 80 nanoseconds at 300 K under NσT at 1 bar.

We simulate each of the eight configurations for an extended period of time to determine if the chains prefer a similar arrangement. Given that the chains are slow to equilibrate at room temperature, the relaxed configurations were simulated at elevated temperature. Based on DSC scans, which show no melting peaks, we simulated the structures at 500K (227 C) under NσT at 1 bar, giving ample time for all conformations to reach a final structure (2000 ns). The structure was then cooled slowly at a rate of 0.4 K/ns to a final temperature of 300K. At elevated temperatures, a weak umbrella potential was applied to the center of mass of chains within a single lamella layer along the π direction to prevent π-stacks from sliding past each other. After cooling, the cell structure and energy were sampled after simulating for a few hundred nanoseconds at room temperature under NVT conditions.

To confirm the crystal structure in solvent cast films, we start from the relaxed π-offset structure and simulate for 2000 nanoseconds at 300 K and 1 bar NσT conditions.

**GIWAXS forward simulation**

GIWAXS forward simulation was carried out using GIWAXSim a custom-built python tool available on github (https://github.com/tchaney97/giwaxsim) operating on the first-order Born approximation to simulate GIWAXS using structure files output from molecular dynamics (MD). To reproduce the orientational disorder found in experimental films, we generated simulated detector images for 16,200 structure orientations. This distribution of structure orientations were selected to simulate the in-plane isotropy inherent in spin-coating, as well as rotation of the polymer aggregates about the backbone direction. A weighted sum of all generated orientations was used to create final GIWAXS simulations, in which the orientation weights were determined using the experimental pole figure shown in **Figure S3**.

The effect of finite aggregate size was simulated by producing rectangular prism slabs of varying sizes with repeated molecular dynamics structure files. The x, y, and z dimensions of the slabs were fit through a gradient descent algorithm, with the loss function set as the squared differences between experimental (chloroform + 5 vol% CN) and simulated GIWAXS patterns. To account for a slight mismatch in q-position of the π-π peak, a radial buffer of 0.1 $Å^{-1}$ was added to better evaluate the fit. Reduced $chi^2$ values were obtained according to the equation below:

$$\chi^2_{reduced} = \frac{\sum_i^N [Resid._i\,]^2}{N - N_{variables}} \qquad \text{eq. 3}$$

Where *N* is the number of pixels, *Resid.* Is the intensity residual between experimental and simulated pixels, and the number of variables, $N_{variables}$ is 3.

**Solution SWAXS**

Solution SWAXS was taken at the Advanced Photon Source on beamline 5-ID-D with an X-ray energy of 17 keV. Three custom Rayonix CCD detectors were used along the evacuated flight tube to collect a wide range of scattering vectors. Data was corrected, azimuthally integrated, and stitched together using the beamline software. Before analysis, empty and neat solvent scattering were subtracted from the solution scattering. Solutions for SWAXS measurement were prepared at a concentration of 10mg polymer per milliliter of solvent. Solutions were allowed to stir at 30°C for >3 hours before measurement. During measurement, solutions were flowed through a custom flow cell consisting of a 0.5 mm OD quartz capillary purchased from Hampton Research that was attached to PTFE tubing and driven by a syringe pump.

**Cryogenic electron microscopy (cryo-EM)**

2D images of PM7 nanocrystals in a chlorobenzene solution were captured using cryo-EM. Quantifoil Holey Carbon Grids, 300 mesh, with 1 μm hole size and 2 μm spacings (Quantifoil MicroTools, Jena, Germany) were used for vitrification of the solution with 10 mg/ml concentration. FEI Vitrobot (FEI Company, Hillsboro, OR) was used for the vitrification of the PM7 solution. The Vitrobot chamber was kept at 4 °C with close to 30% relative humidity. To further reduce the evaporation rate of chlorobenzene during the pipetting and blotting, the Vitrobot chamber was exposed to chlorobenzene vapor by placing 20 ml vials of neat solution in the chamber for 20 minutes before the blotting procedure. The vitrification was done using filter papers mounted to the blotting pads with two cycles of blotting each lasting for 3 seconds, giving a total blotting time of 6 seconds. Due to the solubility of chlorobenzene ice in ethane, liquid nitrogen was used as a coolant liquid for the vitrification of PM7 solution(*64*, *65*). The grids were quickly moved to liquid nitrogen containers after the initial vitrification. High-resolution cryogenic experiments were performed on the FEI Titan Krios at the Huck Institutes of the Life Sciences at the Pennsylvania State University. The measurements were conducted using a 300 kV electron source and a Falcone 4 direct electron detector in linear mode. The nanoprobe mode of the microscope with a 70 μm C2 aperture, with a spot size of 5 was employed to collect the TEM images. All cryo-EM micrographs were collected with an electron dose of $<40$ e/Å$^2$ and an electron rate of $<12$ e/Å$^2$s. Binning of 1 (no camera pixel binning) was used for the images with 4096×4096 pixels. DigitalMicrograph software and ImageJ software were used for fast Fourier transform and licecuts plots of cryo-EM micrographs, respectively(*66*).

# 5. Acknowledgments:

This work was supported by the Office of Naval Research MURI Program Center for Self-Assembled Organic Electronics (SOE), Grant N00014-19-1-2453. T.P.C. acknowledges support from the National Science Foundation Graduate Research Fellowship Program under Grant No. (DGE 2040434). GIWAXS characterization was conducted at beamline 11-BM of the National Synchrotron Light Source II, a U.S. Department of Energy (DOE) Office of Science User Facility operated for the DOE Office of Science by Brookhaven National Laboratory under contract no. DE-SC0012704. SWAXS characterization was conducted at the DuPont-Northwestern-Dow Collaborative Access Team (DND-CAT) located at Sector 5 of the Advanced Photon Source (APS). DND-CAT is supported by Northwestern University, The Dow Chemical Company, and DuPont de Nemours, Inc. This research used resources of the Advanced Photon Source, a U.S. Department of Energy (DOE) Office of Science User Facility operated for the DOE Office of Science by Argonne National Laboratory under Contract No. DE-AC02-06CH11357. Data at DND-CAT was collected using an instrument funded by the National Science Foundation under Award Number 0960140. GIWAXS forward simulation was carried out on the Alpine high performance computing resource at the University of Colorado Boulder. Alpine is jointly funded by the University of Colorado Boulder, the University of Colorado Anschutz, Colorado State University, and the National Science Foundation (award 2201538). The authors acknowledge use of the electron microscopy facilities at the Huck Institutes of the Life Sciences at the Pennsylvania State University.

## 6. Supporting Information:

Supporting Information includes derivations for equations 1-2, structural schematics for polymer stacking, additional GIWAXS analysis, additional figures from molecular dynamics, and additional micrographs from cryo-EM.

## 7. Data Availability:

The data that support the findings of this study are available from the corresponding author upon reasonable request.

## 8. Conflict of Interest:

The authors declare no conflict of interest.

# Supplemental Information for:

# Deciphering the Structure of Push-Pull Conjugated Polymer Aggregates in Solution and Film

Thomas P. Chaney[1], Christine LaPorte Mahajan[2], Masoud Ghasemi[2], Andrew J. Levin[1], Keith P. White[1], Scott T. Milner[2], Enrique D. Gomez[2,3], Michael F. Toney[1,4 *]

[1]*Materials Science and Engineering, University of Colorado, Boulder, CO 80309, USA*

[2]*Department of Chemical Engineering, The Pennsylvania State University, University Park, PA 16802, USA*

[3]*Department of Materials Science and Engineering, The Pennsylvania State University, University Park, PA, 16802, USA*

[4]*Department of Chemical and Biological Engineering and Renewable and Sustainable Energy Institute (RASEI), University of Colorado, Boulder, CO 80309, USA*

**Appendix:** Derivations of equations 1 and 2

The effect of π-offset variations ($\Delta_\pi$) changes on q-position of the (012) in face-on oriented aggregates equation 1:

First, we define the **q** position of the (012) peak using reciprocal space lattice vectors $\boldsymbol{b}^*$ and $\boldsymbol{c}^*$**:**

$$\boldsymbol{q}_{(\mathbf{012})} = \boldsymbol{b}^* + \mathbf{2}\boldsymbol{c}^*$$

The equations for the $\boldsymbol{b}^*$ and $\boldsymbol{c}^*$reciprocal space lattice vectors are defined w.r.t. the real-space lattice vectors ***a, b*** and ***c*** and $V$ the unit cell volume.

$$\boldsymbol{b}^* = \frac{2\pi(\boldsymbol{c}\times\boldsymbol{a})}{V} = \frac{2\pi}{V}\begin{bmatrix} c_y\, a_z - c_z a_y \\ c_z a_x - c_x a_z \\ c_x a_y - c_y a_x \end{bmatrix}$$

$$\boldsymbol{c}^* = \frac{2\pi(\boldsymbol{a}\times\boldsymbol{b})}{V} = \frac{2\pi}{V}\begin{bmatrix} a_y\, b_z - a_z b_y \\ a_z b_x - b_z a_x \\ a_x b_y - a_y b_x \end{bmatrix}$$

Recognizing that changes in the π-offset only alter the $b_x$ component of the real-space unit cell (See **Figure 4 and S16**), we see that $\boldsymbol{b}^*$is unchanged and only the $c_z^*$ component of $\boldsymbol{c}^*$ changes with varying π-offset ($\Delta_\pi$) since $a_z = 0$. The $a_x b_y$ in the $c_z^*$ term also is zero since $b_y = 0$.

$$b_z^* = \frac{2\pi}{V}\left(c_x a_y\right)$$

$$c_z^* = \frac{2\pi}{V}\left(-a_y b_x\right)$$

We can re-define both $b_z^*$ and $c_z^*$ according to easily identified d-spacings from the GIWAXS as well as the length of the monomer. Note the signs of $b_z^*$ and $c_z^*$ are due to the chosen axis system (See **Figures 4 and S16**).

$$b_z^* = \frac{-2\pi}{d_{(010)}}$$

$$c_z^* = \frac{2\pi\Delta_\pi}{d_{(010)} l_{monomer}}$$

Using these definitions of $b_z^*$ and $c_z^*$ we arrive at eq 1:

$$q_{z(012)} = \frac{-2\pi}{d_{(010)}} + \frac{4\pi\Delta_\pi}{d_{(010)} l_{monomer}}$$

The effect of lamella-offset variations ($\Delta_{lamella}$) changes on q-position of the (012) in face-on oriented aggregates equation 2:

We recognize that $\Delta_{lamella}$ only impacts the $a_x$ component of the real-space unit cell (See **Figures 4 and S16**). Connecting this to the definitions of $\boldsymbol{b}^*$ and $\boldsymbol{c}^*$ above, we see that $\boldsymbol{b}^*$ is unchanged and only $c_y^*$ components change with varying $\Delta_{lamella}$ since $c_y = c_z = 0$. Thus, we should expect to see a change in $q_{xy}$ position of the (012) peak with varying $\Delta_{lamella}$.

$$b_x^* = 0, b_y^* = 0$$
$$c_x^* = \frac{2\pi}{V}(a_y b_z)$$
$$c_y^* = \frac{2\pi}{V}(-b_z a_x)$$

We can re-define both $c_x^*$ and $c_y^*$ according to easily identified d-spacings from the GIWAXS as well as the length of the monomer. Note the signs of $c_x^*$ and $c_y^*$ are due to the chosen axis system (See **Figures 4 and S16**). We also find that $b_x^*$=$b_y^*$=0.

$$c_x^* = \frac{-2\pi}{l_{monomer}}$$

$$c_y^* = \frac{-2\pi\Delta_{lamella}}{d_{(100)}l_{monomer}}$$

Using these definitions of $c_x^*$ and $c_y^*$ we arrive at eq 2:

$$q_{xy_{(012)}} = \sqrt{\left(b_x^* + 2c_x^*\right)^2 + \left(b_y^* + 2c_y^*\right)^2}$$
$$= \sqrt{\left(\frac{-4\pi}{l_{monomer}}\right)^2 + \left(\frac{-4\pi\Delta_{lamella}}{d_{(100)}l_{monomer}}\right)^2}$$

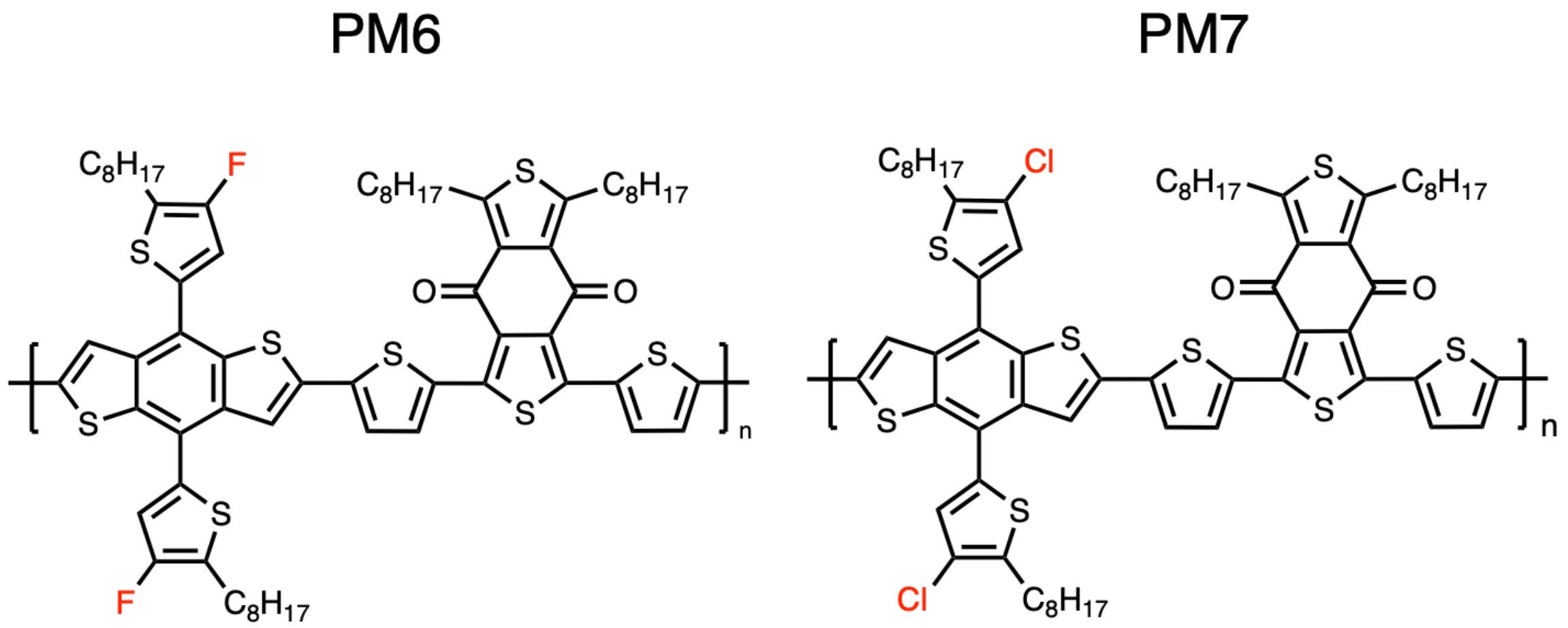

**Figure S1:** Structures of PM6 and PM7 polymers used in this study

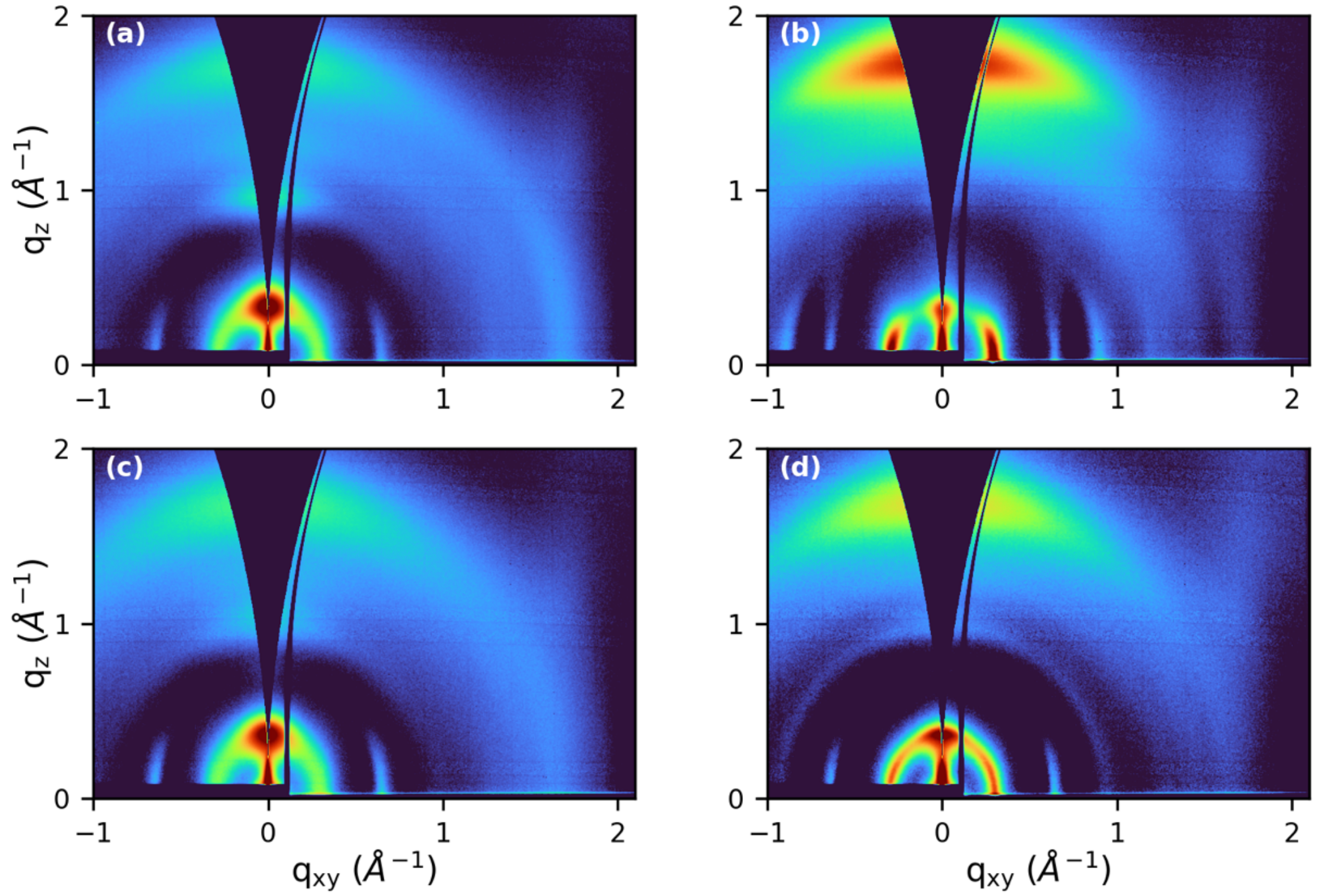

**Figure S2:** Logarithmically scaled GIWAXS detector images converted to q-space are shown for (a) PM6 from neat chloroform neat chloroform, (b) PM6 from chloroform + 5 vol% CN, (c) PM7 from neat chloroform neat chloroform, (d) PM7 from chloroform + 5 vol% CN.

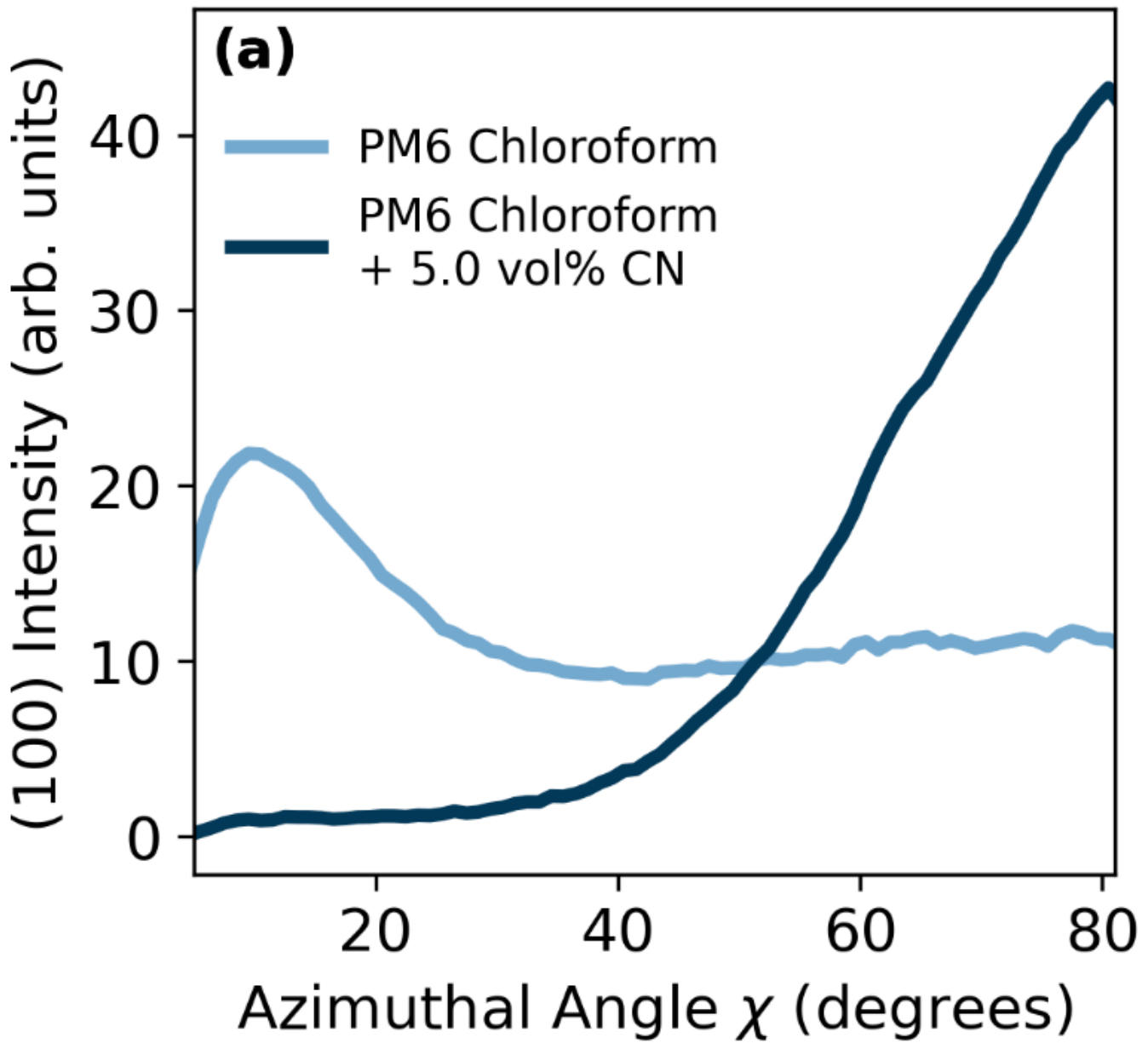

**Figure S3:** Pole figures of lamella (100) integrated peak intensity for PM6 films. Peak intensities are background subtracted and scaled by sin($\chi$) to account for fiber texture of spun coat films (10.1021/la904840q)

**Table S4:** Fitted q-positions and peak widths from GIWAXS

| Film Description | (100) Center [Å$^{-1}$] | (100) FWHM [Å$^{-1}$] | (012) Center [Å$^{-1}$] | (012) FWHM [Å$^{-1}$] | (010) Center [Å$^{-1}$] |
|---|---|---|---|---|---|
| PM6 from CF | 0.304 | 0.126 | 0.646 | 0.060 | 1.740 |
| PM6 from CF + 5% CN | 0.299 | 0.048 | 0.642 | 0.050 | 1.752 |

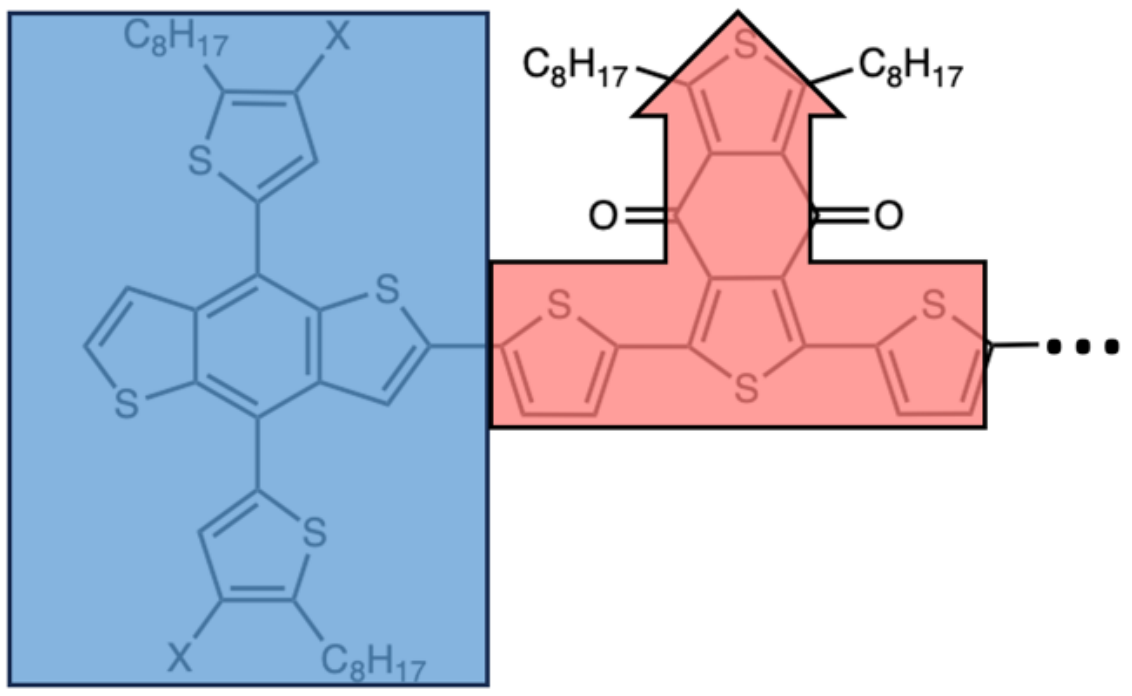

**Figure S5:** Schematic representation of the PM6 monomer.

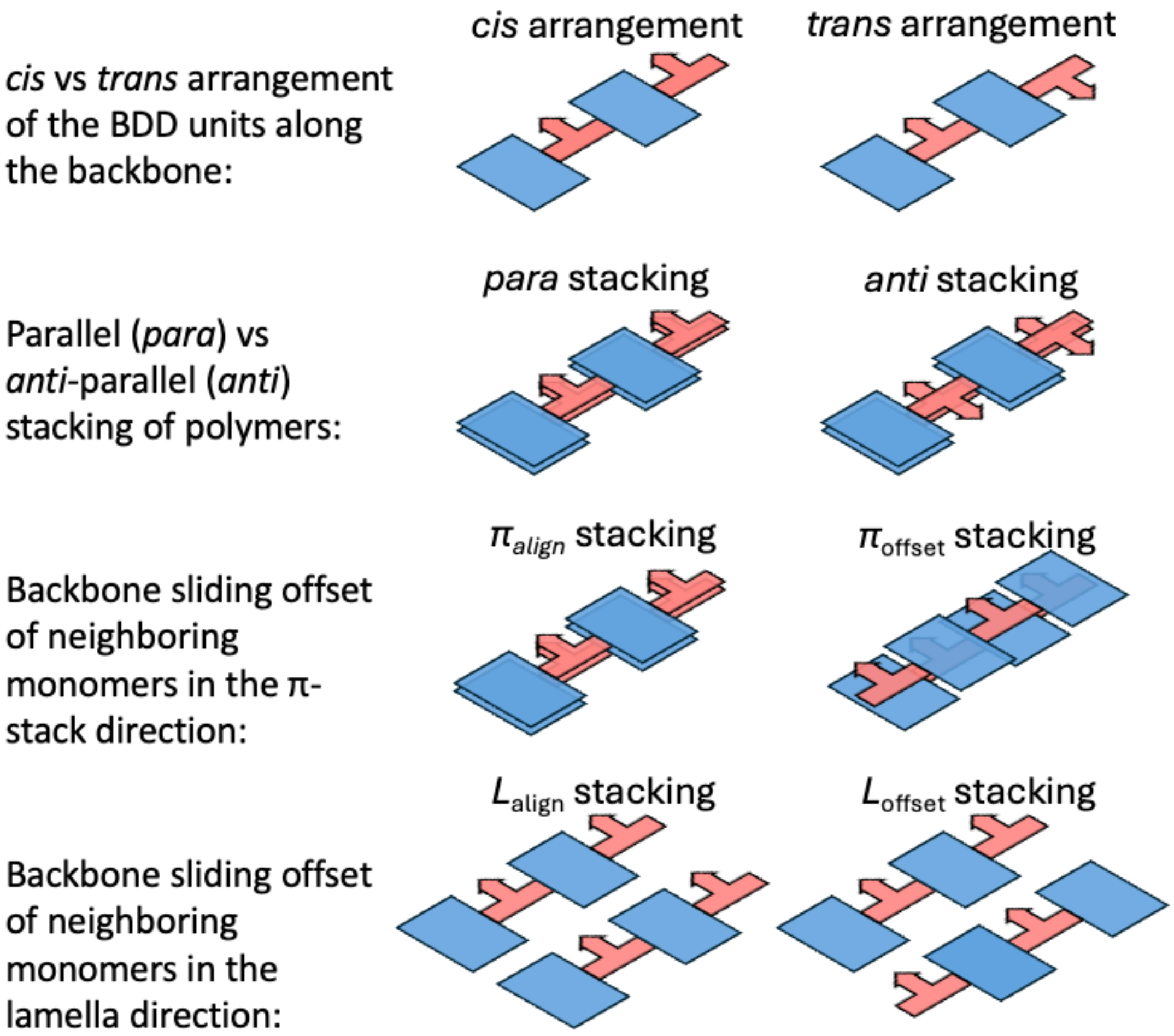

**Figure S6:** Schematic representations of the four PM6 stacking manipulations explored in this study

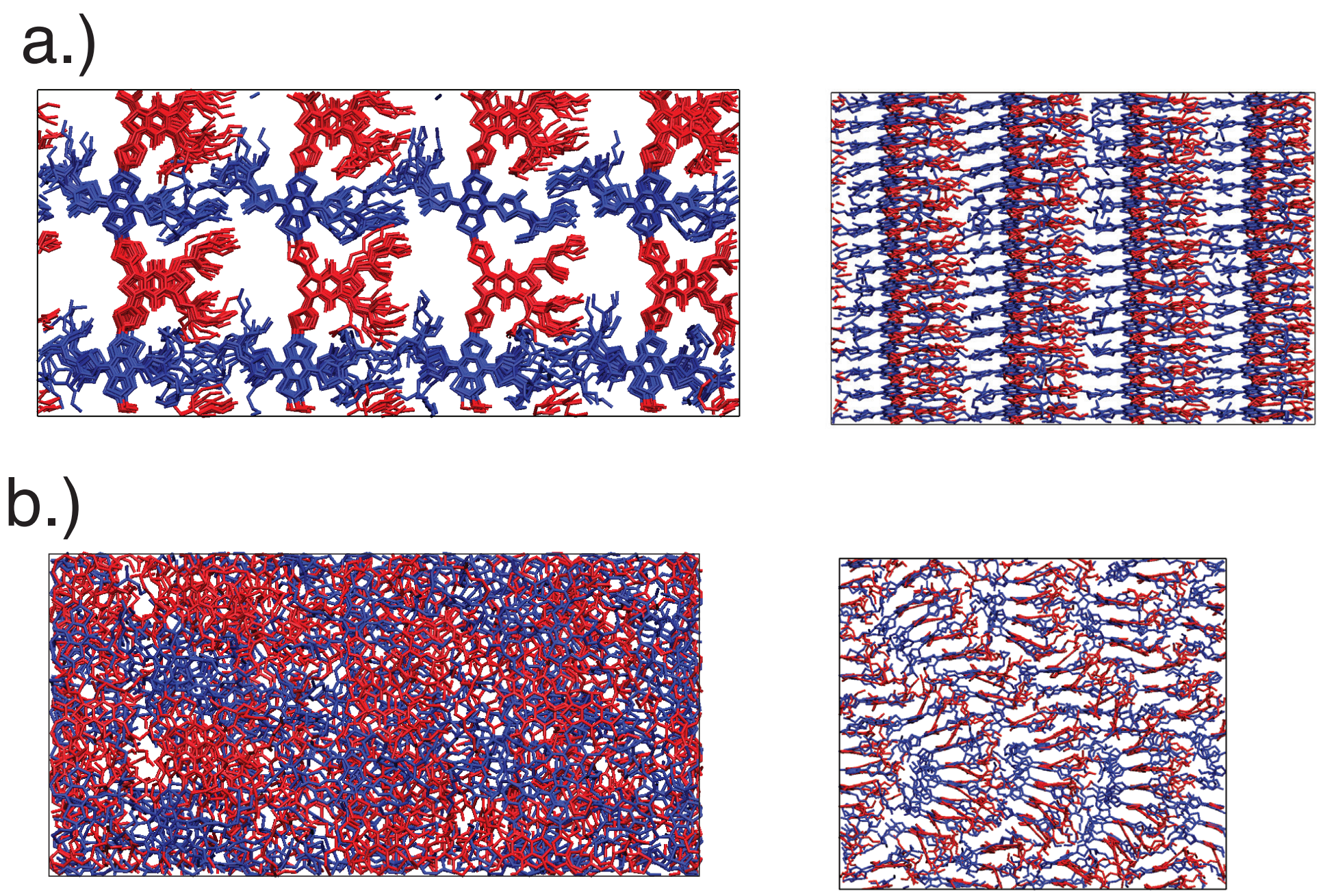

**Figure S7:** Example cis-type aggregate (a) after relaxation with freeze groups and (b) after equilibration without freeze groups. Red bonds correspond to the electron accepting BDD group and blue bonds correspond to the electron donating BDT group. The disappearance of ordered stacking after equilibration in (b) demonstrates the instability of cis-type aggregation.

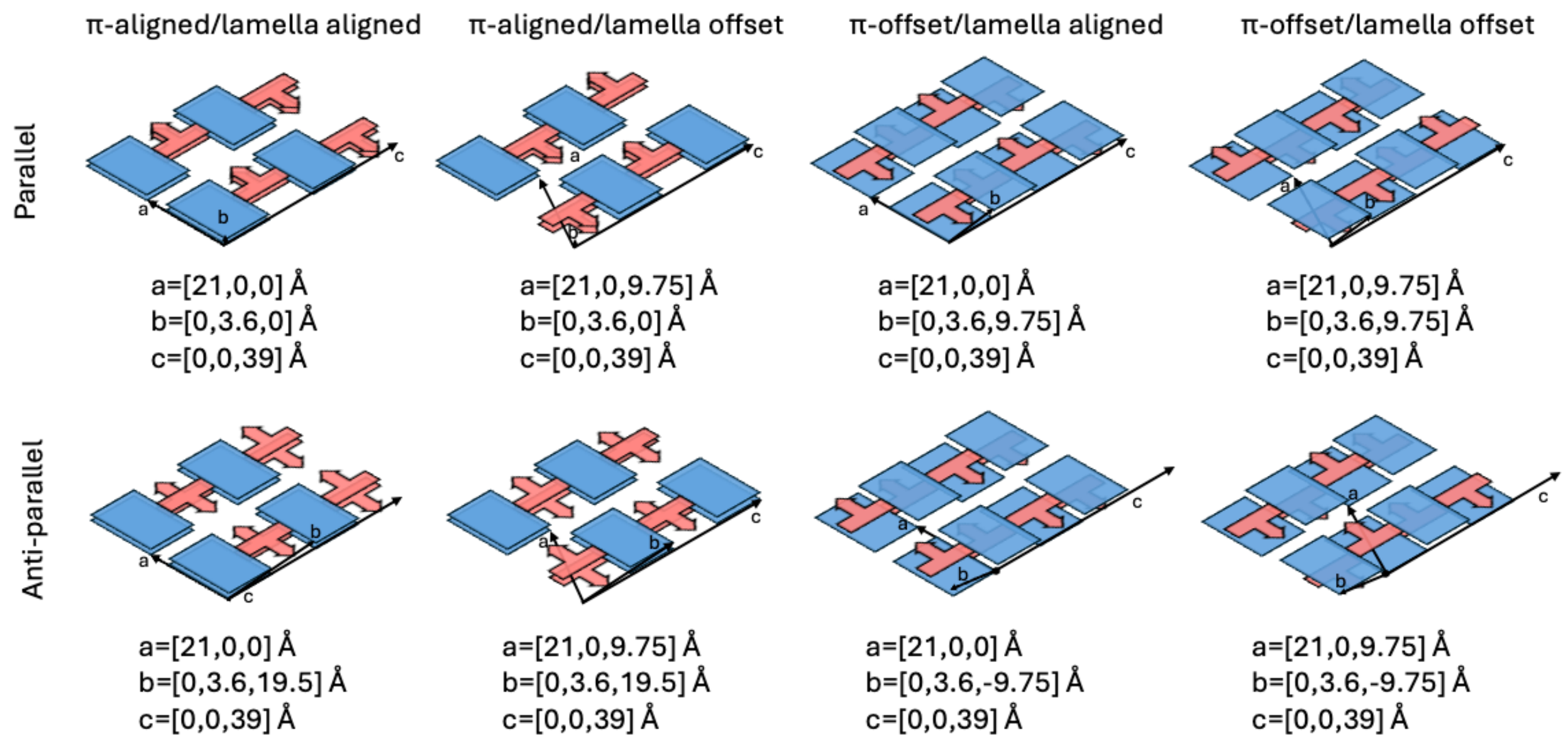

**Figure S8:** Schematic of potential unit cells with trans- backbones considered in this manuscript. Corresponding real-space lattice vectors are shown below each schematic. Color scheme is shown in figure S6.

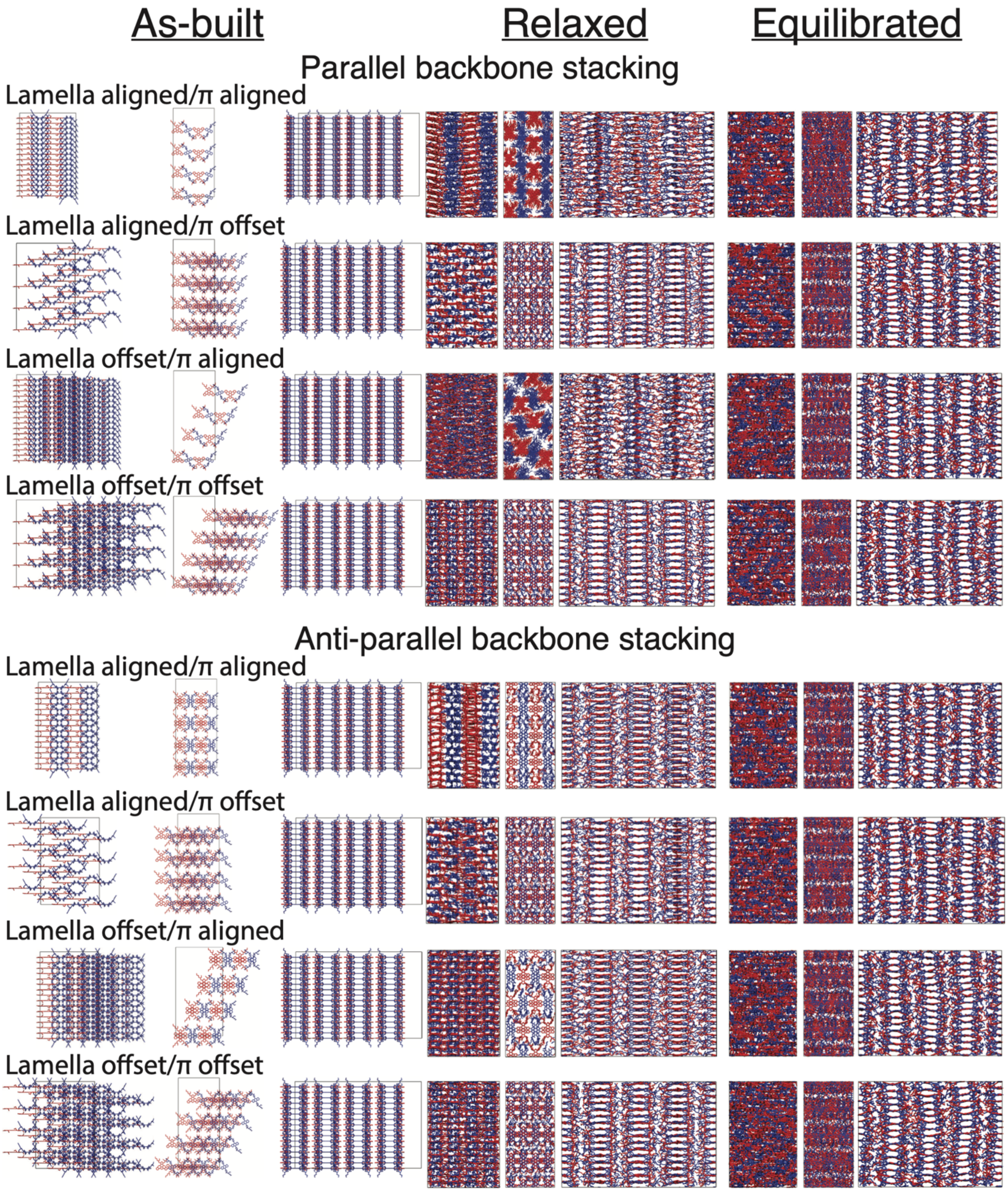

**Figure S9:** Figure showing the configurations as-built, relaxed, and equilibrated using molecular dynamics.

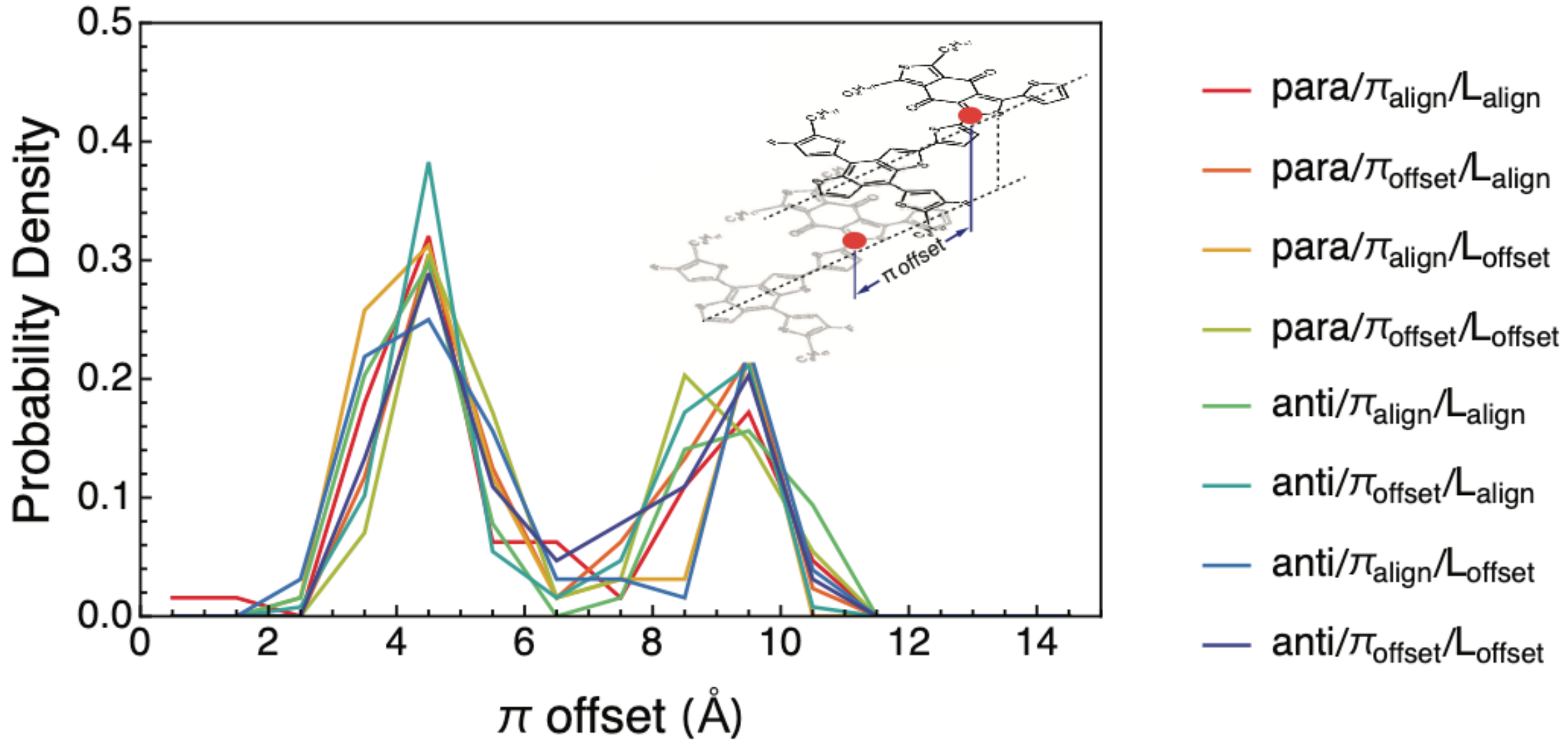

**Figure S10:** Figure showing π offset distribution for all equilibrated configurations.

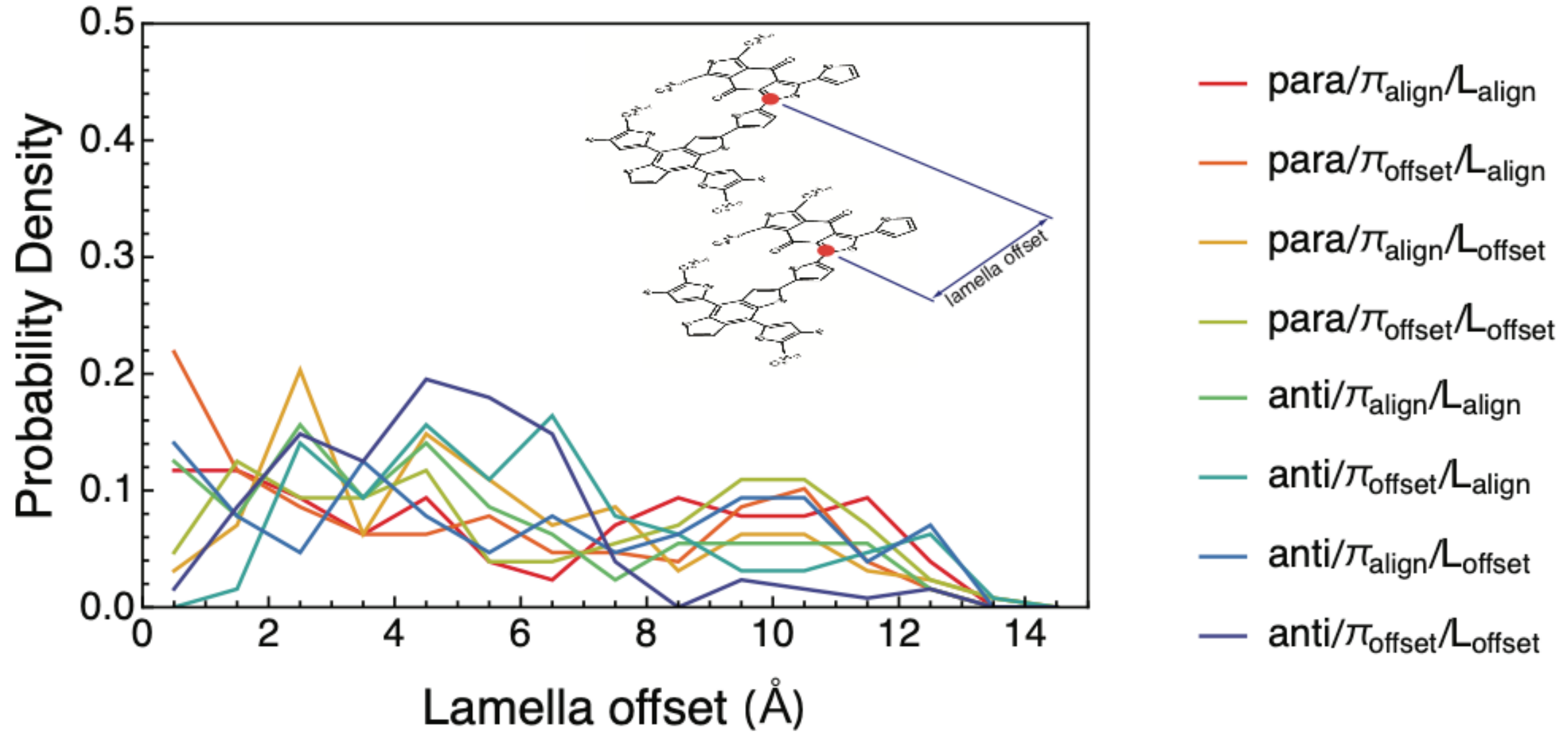

**Figure S11** Figure showing lamella offset distribution for all equilibrated configurations.

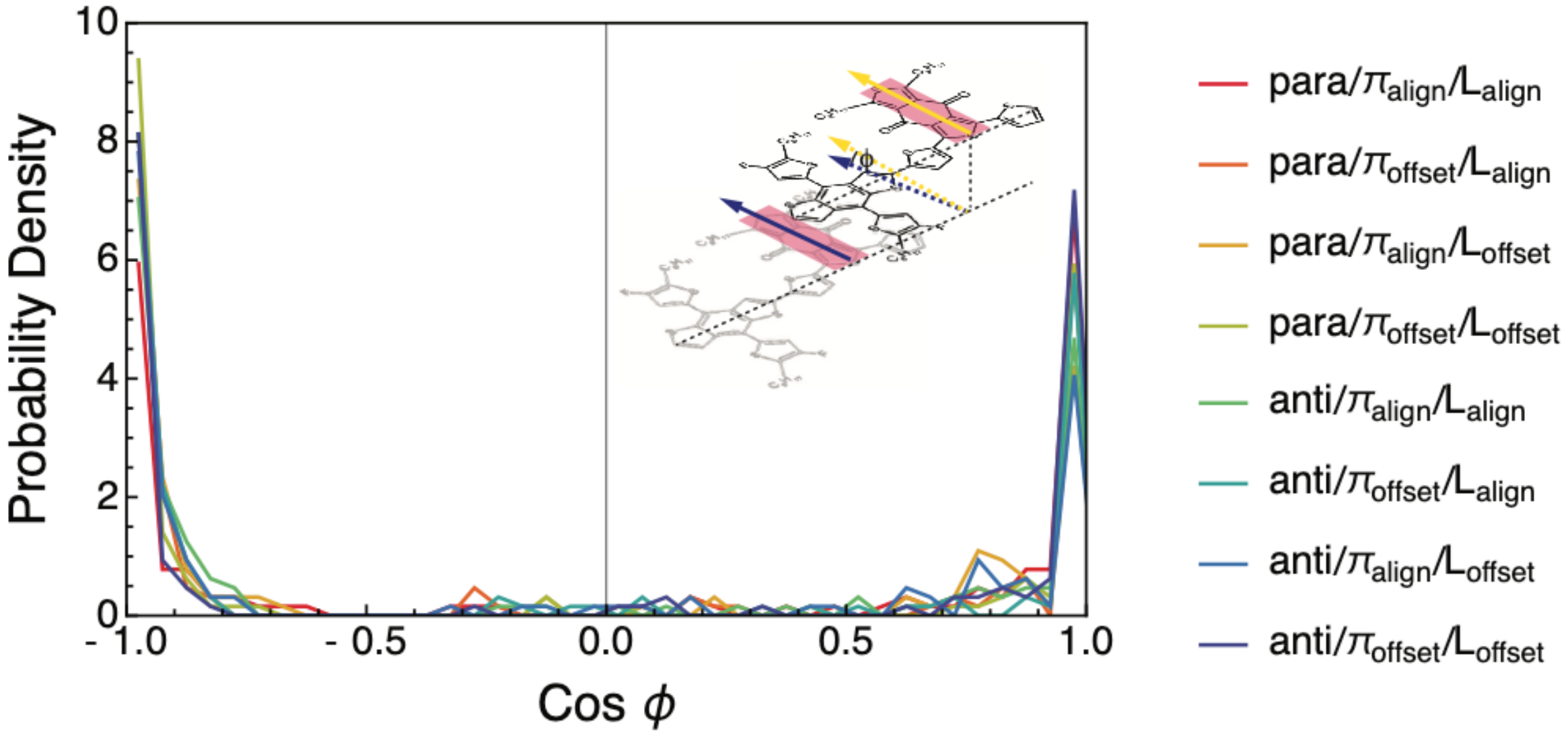

**Figure S12:** Figure showing relative angle between π-stacked dimers

**Table S13** Best fitting slab sizes and reduced chi$^2$ values for each simulation. Fitting details are provided in the methods

| Stacking Motif | Reduced chi$^2$ | Lamella size [Å] | π–π size [Å] | Backbone size [Å] |
|---|---|---|---|---|
| anti/π-aligned/ lamella-aligned | 4648 | 178 | 60 | 98 |
| anti/π-aligned/ lamella-offset | 4647 | 178 | 60 | 98 |
| anti/π-offset/lamella-offset | 4599 | 178 | 69 | 85 |
| anti/π-offset/lamella-aligned | 4839 | 183 | 53 | 100 |
| parallel/π-aligned/lamella-aligned | 8765 | 193 | 49 | 92 |
| parallel/π-aligned/lamella-offset | 4285 | 176 | 61 | 94 |
| parallel/π-offset/lamella-offset | 5140 | 177 | 54 | 92 |
| parallel/π-offset/lamella-aligned | 4689 | 179 | 57 | 94 |

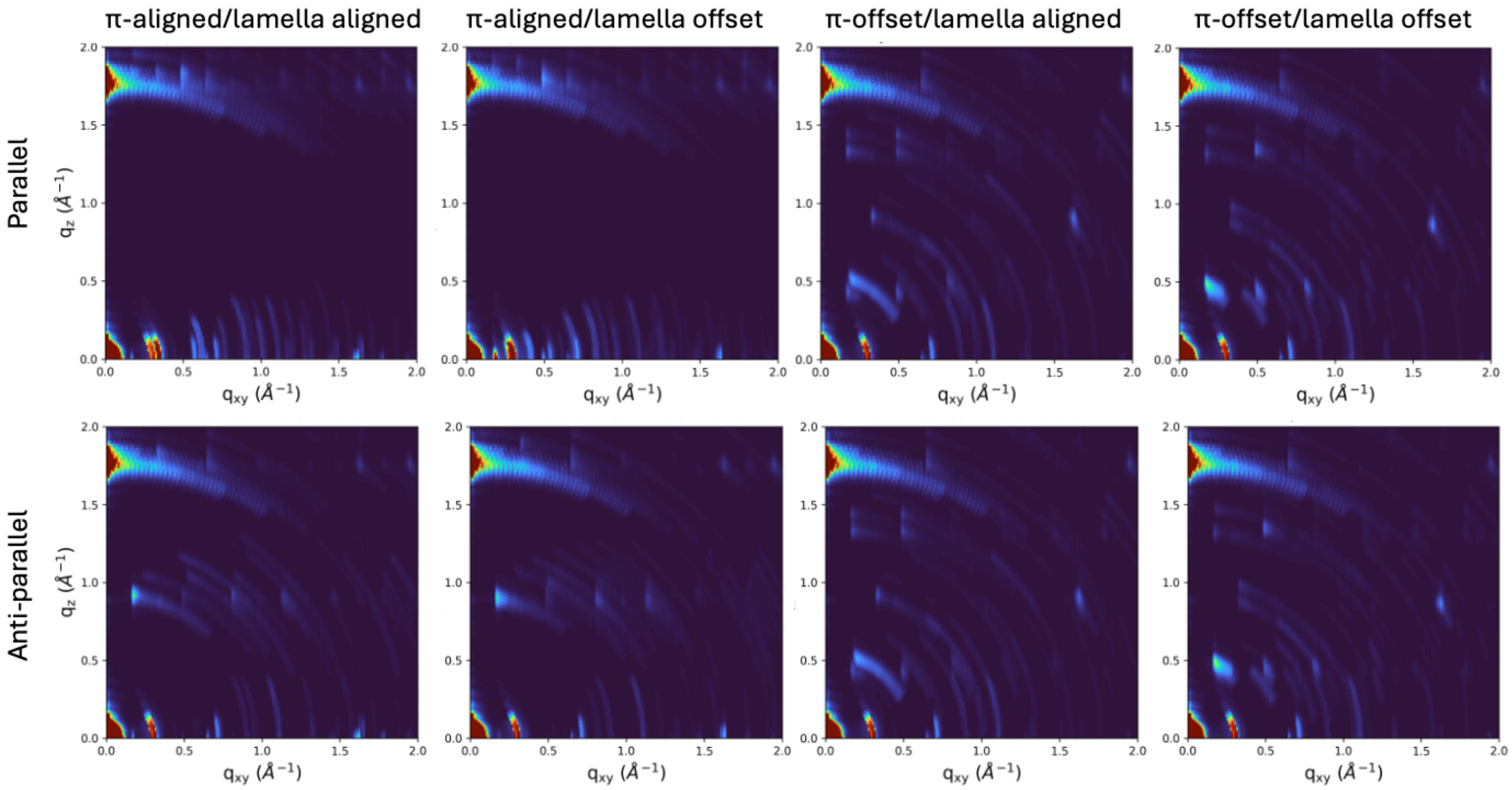

**Figure S14** Simulated GIWAXS q-space images (linearly scaled intensities) of PM6 arranged in all the 8 trans configurations shown in figure S4 before any relaxation. All unit cells are oriented face-on. Fiber texture and rotational disorder about the backbone axis is incorporated. A slab size of 300x80x300 Å was used for all simulations shown.

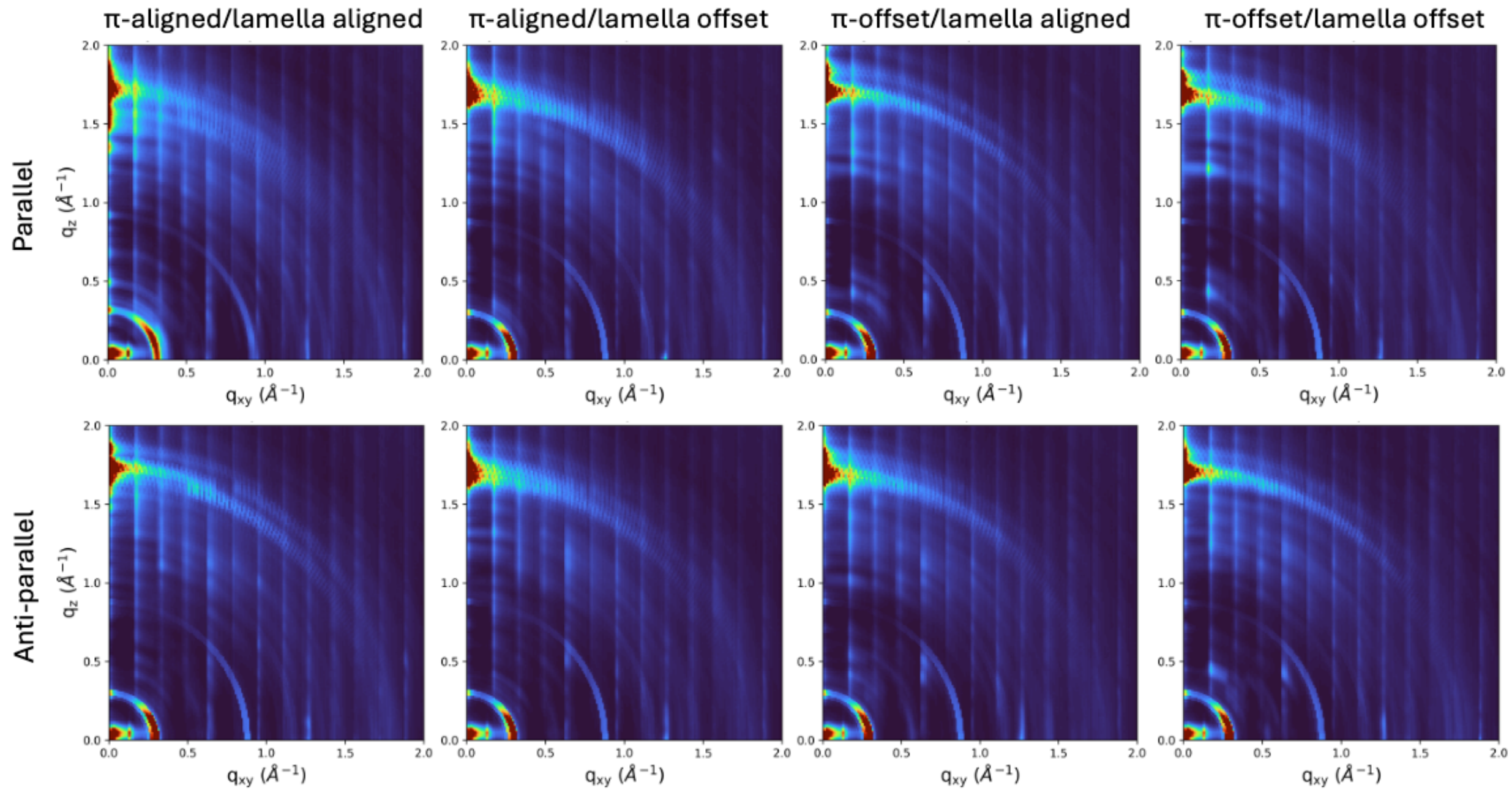

**Figure S15:** Simulated GIWAXS q-space images (linearly scaled intensities) of PM6 arranged in all the 8 trans configurations shown in figure S4 after equilibration. All unit cells are oriented face-on. Fiber texture and rotational disorder about the backbone axis is incorporated. Vertical streaks are artifacts from the perfectly repeating trans configuration resulting in Bragg peaks with 0.15$Å^{-1}$ spacings. A slab size of 300x80x300 Å was used for all simulations shown.

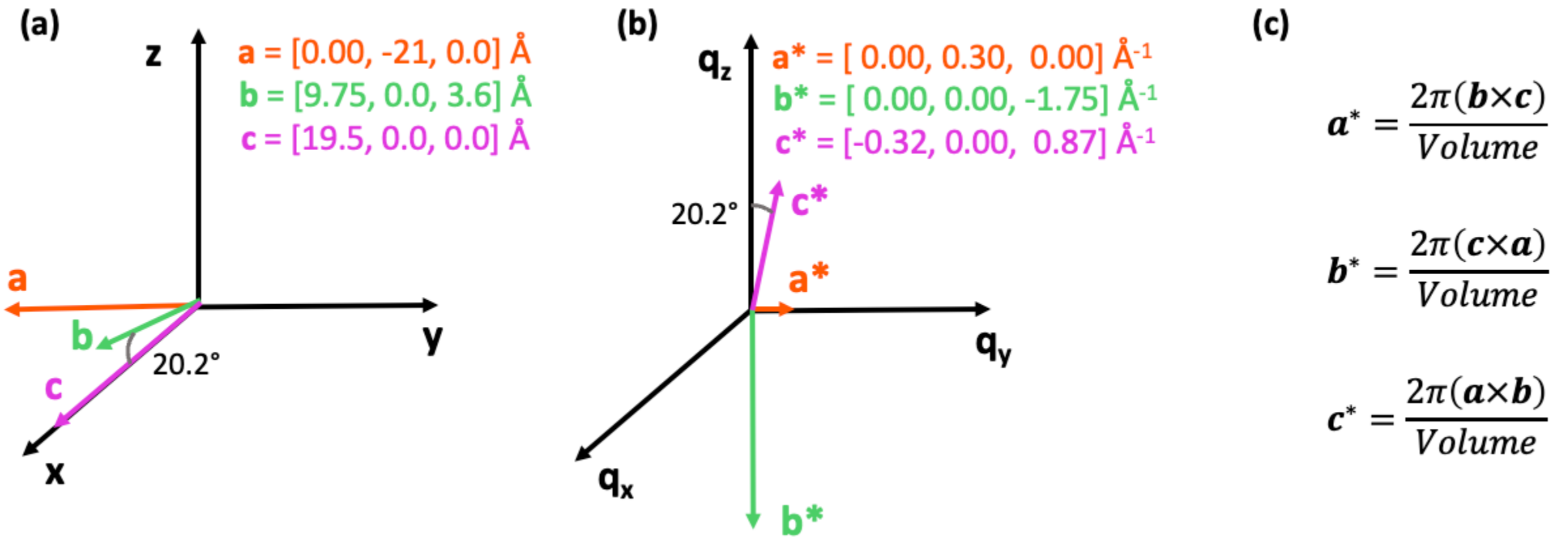

**Figure S16** (a) Proposed monoclinic real space unit cell and (b) the associated reciprocal space unit cell. (c) Describes the equations that relate the real-space lattice to the reciprocal space lattice where *Volume* is the real-space unit cell volume

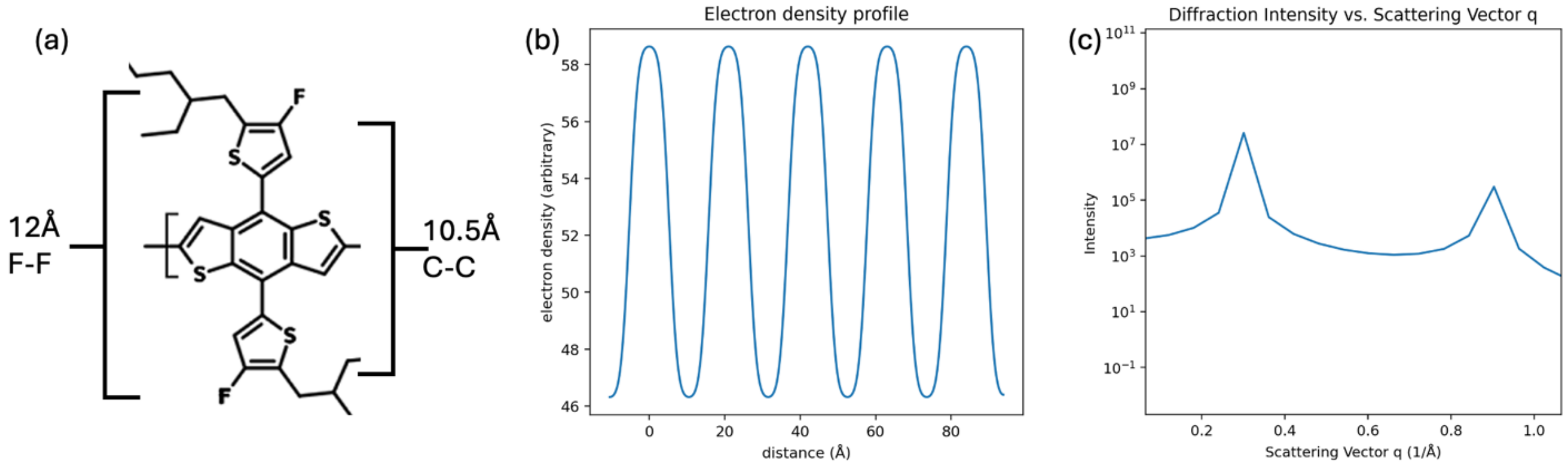

**Figure S17:** Visualization and simulation of the suppression of (200) peak due to structure factor. (a) dimensions of BDT unit showing a conjugated length of nearly exactly half that of the lamella spacing. (b) smoothed electron density profile where regions of low electron density (alkyl chain regions) are the same width as high electron density regions (conjugated BDT unit). (c) norm squared of the FFT on the electron density profile showing the presence of a (100) and (300) at 0.3 and 0.9 $Å^{-1}$ with no signal at 0.6 $Å^{-1}$ where a (200) would be expected.

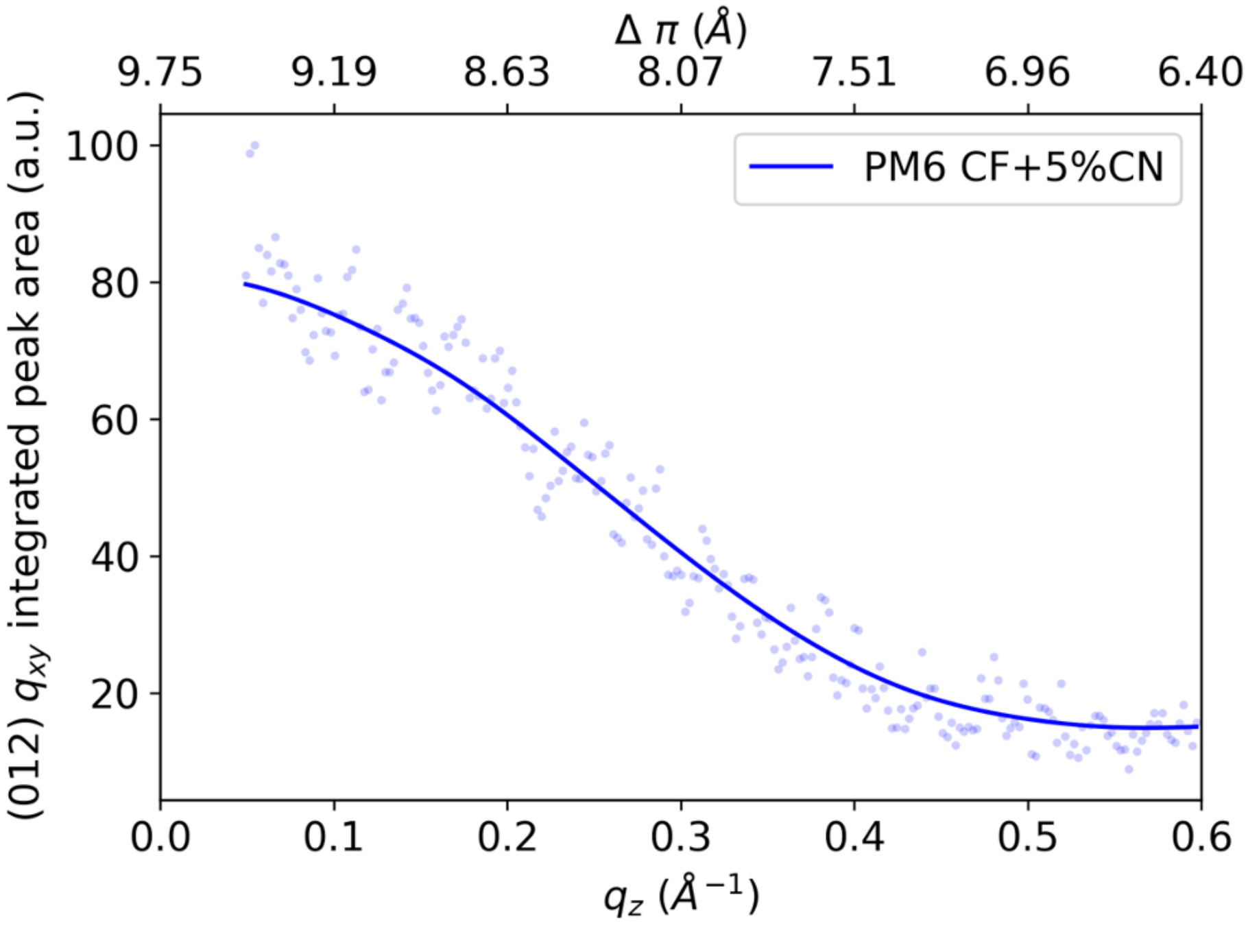

**Figure S18:** Analysis of $q_z$ distribution of (012) backbone peak. Offset between π-stacks was computed as described in equation 1. Note that $\Delta_\pi$ values here are decreasing with increasing $q_z$ as values of $\Delta_\pi$ greater than ½ $l_{monomer}$ are converted to the smallest absolute offset given by $l_{monomer} - \Delta_\pi$. The solid line represents smoothed average of data. Note the data cuts off before $q_z = 0$ to avoid capturing Yoneda peak intensity.

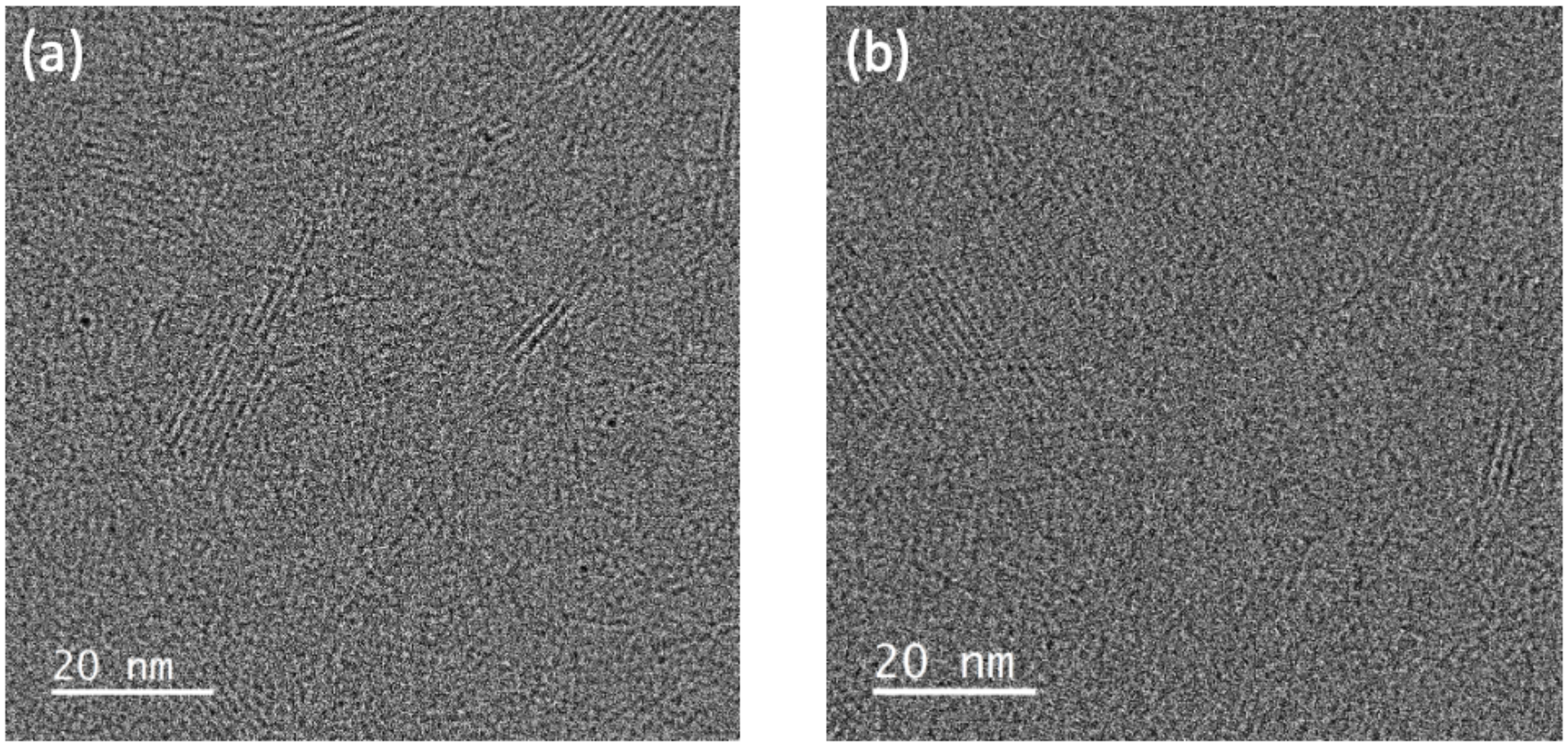

**Figure S19:** Addition cryo-EM micrographs of vitrified PM7 in chlorobenzene solutions